\documentclass[twocolumn,showpacs,twoside,10pt,aps,pra,superscriptaddress]{revtex4-2}
\usepackage{amsmath}
\usepackage{float}
\usepackage{graphicx}
\usepackage{dcolumn}
\usepackage{bm}
\usepackage{url}
\usepackage[colorlinks=true,linkcolor=blue, anchorcolor=blue, citecolor=blue]{hyperref}
\usepackage{orcidlink}
\usepackage{braket}
\usepackage{appendix}
\usepackage{ulem}
\usepackage{enumerate}
\usepackage{multirow}
\usepackage{cases}
\usepackage{amsthm,amsmath,amssymb}
\usepackage{mathrsfs}
\usepackage{tabularx}
\usepackage{booktabs} 
\usepackage[ruled,vlined]{algorithm2e}
\SetKwProg{For}{for}{}{}
\SetKwComment{Comment}{$\triangleright$\ }{}
\usepackage{algpseudocode}
\SetKwInOut{Require}{Input}
\SetKwInOut{Ensure}{Output}

\begin{document}
\preprint{APS/123-QED}

\title{Credible-interval-based adaptive Bayesian quantum frequency estimation for entanglement-enhanced atomic clocks}
\def\SYSUZH{Laboratory of Quantum Engineering and Quantum Metrology, School of Physics and Astronomy,\\ 
Sun Yat-Sen University (Zhuhai Campus), Zhuhai 519082, China}
\def\SZU{Institute of Quantum Precision Measurement, State Key Laboratory of Radio Frequency Heterogeneous Integration,\\
College of Physics and Optoelectronic Engineering, Shenzhen University, Shenzhen 518060, China}
\def\GD{Quantum Science Center of Guangdong-Hong Kong-Macao Greater Bay Area (Guangdong), Shenzhen 518045, China}

\author{Jungeng Zhou~\orcidlink{0009-0004-1925-2555}}
  \affiliation{\SYSUZH}
   \affiliation{\SZU}

\author{Jiahao Huang~\orcidlink{0000-0001-7288-9724}}
  \affiliation{\SYSUZH}
   \affiliation{\SZU}   
   
\author{Jinye Wei~\orcidlink{0009-0001-0965-8058}}
  \affiliation{\SYSUZH}
   \affiliation{\SZU}

\author{Chengyin Han~\orcidlink{0000-0003-0771-8819}}
  \affiliation{\SZU}
  \affiliation{\GD}
  
\author{Chaohong Lee~\orcidlink{0000-0001-9883-5900}}
  \altaffiliation{Email: chleecn@szu.edu.cn, chleecn@gmail.com}
  \affiliation{\SZU}
  \affiliation{\GD}
  
\begin{abstract}
Entanglement-enhanced quantum sensors encounter a fundamental trade-off: while entanglement improves precision to the Heisenberg limit, it restricts dynamic range.
To address this trade-off, we present a credible-interval-based adaptive Bayesian quantum frequency estimation protocol for Greenberger-Horne-Zeilinger (GHZ)-state-based atomic clocks.
Our method optimally integrates prior knowledge with new measurements and determines the interrogation time by correlating it with the period of the likelihood function, based on Bayesian credible intervals.
Our protocol can be implemented using either individual or cascaded GHZ states, thereby extending the dynamic range without compromising Heisenberg-limited sensitivity.
In parallel with the cascaded-GHZ-state protocol using fixed interrogation times, the dynamic range can be extended through an interferometry sequence that employs individual GHZ states with variable interrogation times.
Furthermore, by varying the interrogation times, the dynamic range of the cascaded-GHZ-state protocol can be further extended.
Crucially, our protocol enables dual Heisenberg-limited precision scaling $\propto 1/(Nt)$ in both particle number $N$ and total interrogation time $t$, surpassing the hybrid scaling $\propto 1/{(N\sqrt {t}})$ of the conventional cascaded-GHZ-state protocol.
While offering a wider dynamic range, the protocol is more stable against noise and more robust to dephasing than existing adaptive schemes.
Beyond atomic clocks, our approach establishes a general framework for developing entanglement-enhanced quantum sensors that simultaneously achieve both high precision and broad dynamic range.
\end{abstract}

\maketitle


\section{Introduction}
Multi-particle quantum entanglement is a key resource for achieving the fundamental precision limit in quantum sensing~\cite{degen2017quantum,ye2024essay,huang2024entanglementenhanced,zhuang_Quantum_2024}. 
For a probe of $N$ uncorrelated particles, the measurement precision can only reach the standard quantum limit (SQL) with a scaling of $1/\sqrt{N}$~\cite{giovannetti2004quantumenhanced}. 
By employing multi-particle entangled states~\cite{giovannetti2006quantum,nonclassicalstates,lee_Adiabatic_2006,giovannetti2011advances}, the measurement precision can surpass the SQL. 
In particular, the Greenberger-Horne-Zeilinger (GHZ) state~\cite{Greenberger1989} can achieve the Heisenberg limit with a scaling of $1/N$~\cite{Heitler1954}.
Atomic clocks~\cite{ludlow2015optical}, the most accurate and precise device for time-keeping, are rapidly emerging as a significant area of interest in entanglement-enhanced quantum metrology. 
Entanglement-enhanced atomic clocks~\cite{colombo2022entanglementenhanceda}, utilizing spin-squeezed states~\cite{schulte2020prospects,appel2009mesoscopica,leroux2010orientationdependenta,pedrozo-penafiel2020entanglement,eckner2023realizing} and GHZ states~\cite{cao2024multiqubit,finkelstein2024universal,kielinski2024ghz}, are essential for advanced science and technology, with extensive applications in timekeeping, navigation, astronomy, and space exploration~\cite{chou2010opticala,kennedy2020precision,schkolnik2023optical}.

Based on frequentist measurements of individual entangled states, entanglement-enhanced atomic clocks cannot simultaneously achieve both high precision and high dynamic range~\cite{degen2017quantum}.
For example, although leveraging GHZ states can yield a $\sqrt{N}$-fold improvement in precision compared to non-entangled states, the corresponding dynamic range is reduced by a factor of $N$.
This is a result of frequency amplification, which narrows the interval width from $1/T$ to $1/(NT)$ with $T$ representing the Ramsey interrogation time.
This issue can be effectively addressed by employing a sequence of NOON states~\cite{higgins2007entanglementfree,berry2009how,higgins2009demonstrating,kessler2014heisenberglimited} or a cascade of GHZ states with exponentially increasing particle numbers~\cite{komar2014quantum,liu2023fullperiod,cao2024multiqubit,finkelstein2024universal}. 
These approaches facilitate the update of probability distributions through Bayesian estimation, as the overlap of likelihood functions across different periods mitigates phase ambiguity.
Recently, entanglement-enhanced atomic optical clocks utilizing cascaded GHZ states have been demonstrated with optical tweezer arrays~\cite{cao2024multiqubit,finkelstein2024universal}.

In addition to using cascaded GHZ states, the dynamic range can be extended by using Bayesian measurements with different interrogation time.
By combining interferometry measurements with short and long interrogation times~\cite{borregaard2013efficient,han2024atomic,ma2025adaptive,wei2025adaptive,wei2025practical}, one can achieve a high dynamic range and improve sensitivity through adaptive Bayesian quantum estimation.
In most existing adaptive protocols~\cite{said_Nanoscale_2011,cappellaro2012spinbath,bonato2016optimized,Craigie2021}, the adaptive Bayesian quantum estimation procedure is implemented using an interferometry sequence with exponentially varying interrogation times.
By adjusting the auxiliary phase derived from the Fourier coefficients of the posterior distribution, one can minimize the Holevo variance~\cite{holevo1984covarianta}, which typically requires a linear change in the number of copies as the interrogation time varies.
However, implementing these Bayesian approaches using cascaded GHZ states proves to be highly challenging.
This problem may be resolved if the copies of GHZ states with different particle numbers remain fixed during the adaptive process.

In this article, we demonstrate how to utilize Bayesian quantum estimation to extend the dynamic range of GHZ-state-based atomic clocks while achieving dual Heisenberg scaling of precision. 
In the context of Bayesian quantum estimation, we design a sequence of correlated Ramsey interferometry for atomic clocks with individual or cascaded GHZ states.  
We have developed an adaptive Bayesian quantum frequency estimation protocol based on the credible interval, named the credible-interval-based adaptive Bayesian quantum frequency estimation (denoted as ``CI-adaptive" protocol for short).
Our protocol employs an interferometry sequence with increasing interrogation time, which ensures the period of the next likelihood function corresponds to the credible interval width of the current posterior distribution.
Compared to the cascaded-GHZ-state protocol~\cite{komar2014quantum,liu2023fullperiod,cao2024multiqubit,finkelstein2024universal} with fixed interrogation time $T_{\text{max}}$, our method can achieve a comparable or superior dynamic range by a sequence of interferometry measurements with variable interrogation time (ranging from $T_{\text{min}}$ to $T_{\text{max}}$) using only an individual GHZ state. 
By integrating cascaded GHZ states with variable interrogation times, the dynamic range can be further improved by a factor of $T_{\text{max}}/T_{\text{min}}$.

Moreover, our protocol enables for a stable expansion of the dynamic range with few copies.
By fixing the number of measurement copies $M$, this can be achieved by selecting a sufficiently high credible level $\zeta$ (subject to the condition of the scaling factor $\alpha\ge1$).
The CI-adaptive protocol demonstrates significantly improved resilience to experimental noise and remains robust under realistic dephasing.
It surpasses existing adaptive strategies, such as the Fourier-coefficients-based adaptive Bayesian protocol~\cite{bonato2016optimized,Craigie2021} (denoted as ``FC-adaptive'' protocol for short), in both precision scaling and dynamic range.
Our approach thus enables adaptive Bayesian protocols using both individual and cascaded GHZ states, making it particularly well-suited for atomic arrays capable of simultaneous preparation and measurement~\cite{cao2024multiqubit,finkelstein2024universal}.
Notably, surpassing the hybrid scaling $\propto 1/(N\sqrt{t})$, our scheme achieves a dual Heisenberg scaling $\propto 1/(Nt)$ in both particle number $N$ and total interrogation time $t$.
In our protocol, the integration of noise resilience, high dynamic range, and improved measurement precision offers significant advantages for the development of next-generation entanglement-enhanced quantum sensors.

\begin{figure*}[tbh]
    \centering
    \includegraphics[width=\linewidth]{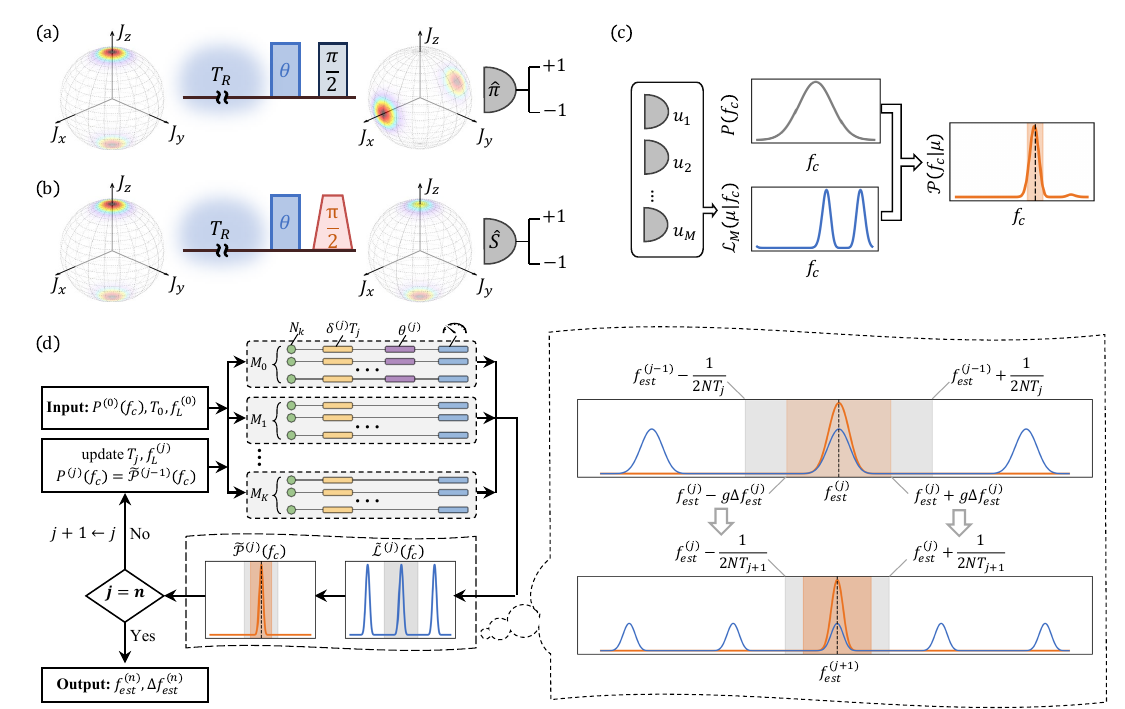}
    \caption{{\bf Schematics of GHZ-state-based Ramsey interferometry and credible-interval-based adaptive Bayesian estimation.} 
    (a) Ramsey interferometry with a GHZ state using parity measurement. 
    (b) Ramsey interferometry with a GHZ state using one-axis twisting interaction-based readout. 
    (c) A Bayesian estimation process demonstrating likelihood calculation based on measurement outcomes and subsequent posterior distribution updating. 
    (d) Left: Implementation of the CI-adaptive protocol with $(K+1)$ ensembles of cascaded GHZ states.
    Each group (labeled by $k$) includes $M_k$ copies of $N_k$-particle GHZ states. 
    The auxiliary phase $\theta$ is applied only to the group with $k=0$. 
    The likelihood distribution $\tilde{\mathcal{L}}^{(j)} (f_c)$ is obtained via measurement and then is used to update the posterior distribution $\tilde{\mathcal{P}}^{(j)} (f_c)$.
    The current posterior distribution $\tilde{\mathcal{P}}^{(j)} (f_c)$ is then set as the next prior distribution $P^{(j+1)} (f_c)$ and gives the next interrogation time $T_{j+1}$. 
    Right: Credible-interval update. 
    The next interrogation time $T_{j+1}$ is determined by the credible interval (orange region) of the current posterior distribution (orange curve). 
    This constrains the likelihood period (gray region) to match twice the credible interval width, expressed as $1/(N_0 T_{j+1}) = 2g \Delta f_{\text{est}}^{(j)}$.}
    \label{fig:1}
\end{figure*}

\section{Model \label{sec:Model}}
\noindent
\subsection{GHZ-state-based Bayesian quantum frequency estimation \label{sec:GHZ}}

We consider Bayesian frequency estimation based on GHZ states in an ensemble of $N$ identical two-level particles.
The system can be described by a collective spin $\bm{\hat{J}}=\{\hat{J}_x,\hat{J}_y,\hat{J}_z\}$ with $\hat{J}_{\alpha}=\frac{1}{2}\sum_{l=1}^{N}{\hat{\sigma}_{\alpha}^{(l)}}$ and $\hat{\sigma}^{(l)}_{\alpha}$ being the Pauli matrix of the $l$-th particle~\cite{biedenharn1984angular}. 
Given an input GHZ state $\ket{\psi}_{GHZ} =  (\ket{0}^{\otimes N} + \ket{1}^{\otimes N} )/\sqrt{2}$,
it will collectively accumulate a relative phase $\phi = \delta T$ during signal interrogation governed by the Hamiltonian $\hat{H}/\hbar = \delta \hat{J}_z$ (with $\hbar=1$ hereafter).
Here $\delta = 2\pi (f_c - f_L)$ is the detuning between the local oscillator (LO) frequency $f_L$ and the clock transition frequency $f_c$ between $\ket{0}$ and $\ket{1}$, and $T$ is the interrogation time.
The LO generates a periodic frequency signal that is locked to the atoms, in which its frequency is repeatedly referenced to an atomic transition frequency by monitoring the atomic response and applying a feedback correction. 
Thus, the output state reads $(\ket{0}^{\otimes N} + e^{i N \phi}\ket{1}^{\otimes N} )/\sqrt{2}$. 
Meanwhile, one may vary the accumulated phase by introducing an auxiliary phase $\theta = 2\pi N f_a T$, which can be realized by applying a frequency shift $f_a$ to the LO frequency~\cite{liu2023fullperiod,finkelstein2024universal,shaw2024multiensemblea}.
Finally, one can perform parity measurement~\cite{bollinger1996optimal,campos2003optical,qian2005universal,huang2015quantum} or interaction-based readout~\cite{colombo2022timereversalbased,davis2016approaching,nolan2017optimal,huang2018achieving,ma2024phase} to obtain $f_c$.

For both parity measurement and interaction-based readout [see Figs.~\ref{fig:1}~(a) and (b) and Supplementary Information], the conditional probability of obtaining $u$ can be written in a unified form, 
\begin{eqnarray}
    \label{eq:p_sin}
    \mathcal{L}_{u} &\equiv& \mathcal{L}(u|N,\theta,f_c,f_L,T) \nonumber \\
    &=& \frac{1}{2}\{1 + u  \xi(N) C \sin[2\pi NT (f_c-f_L)+\theta]\},
\end{eqnarray}
where $u=\pm 1 $ denote the even/odd outcomes for parity measurement $\hat{\Pi} = e^{i\pi \lfloor \hat{J}_z \rfloor}$ or the positive/non-positive outcomes for sign measurement $\hat{S} = \textrm{Sgn} [\hat{J}_z]$ used in interaction-based readout, $\xi(N)=\pm 1$ depends on $N$ and the measurement, and the contrast $C$ depends on the dephasing and detection noise under realistic experimental conditions.

Bayesian quantum estimation provides a powerful tool for optimal high-precision measurements~\cite{pezze2014quantum,li2018frequentist}.
For a given initial prior distribution $P(f_c)$, where $f_c$ is the unknown frequency to be estimated, knowledge about $f_c$ is updated after each measurement using Bayes' theorem.
The posterior distribution obtained at each step can be iteratively used as the prior for the subsequent step, enabling sequential Bayesian updating.
As shown in Fig.~\ref{fig:1}(c), after $M$ independent measurements (or equivalently, simultaneous measurement of $M$ copies), the final posterior distribution is given by
\begin{equation}\label{eq:PM}
    \mathcal{P}_{M} (f_c|N,\theta,f_L,T,\mu) =\mathcal{N} \mathcal{L}_{M}(\mu|N,\theta,f_c,f_L,T)P(f_c),
\end{equation}
where $\mathcal{L}_{M}(\mu|N,\theta,f_c,f_L,T) = \binom{M}{\mu} { [\mathcal{L}_{+}]^{\mu} [\mathcal{L}_{-}]^{M-\mu}}$ is the likelihood function for the measurement sequence, with the result $u=+1$ occurring $\mu$ times and $u=-1$ occurring $M-\mu$ times. Here, $\mathcal{N}$ is the normalization constant.

As shown in Fig.~\ref{fig:1}~(d), we consider the cascaded GHZ states comprising $K+1$ ensembles from $k=0$ to $k=K$, in which the $k$-th ensemble contains $M_k$ copies of a $N_k$-qubit GHZ state~\cite{berry2009how,cao2024multiqubit,finkelstein2024universal}. 
The likelihood function then reads 
\begin{equation}
    \label{eq:L_total} 
    \tilde{\mathcal{L}}(f_c) = \prod_{k=0}^{K}{\mathcal{L}_{M_k}(\mu_k|N_k,\theta_k,T,f_c)},
\end{equation}
where $\mathcal{L}_{M_k}(\mu_k|N_k,\theta_k,f_c,f_L,T) $ is the likelihood function of the individual ensemble as in Eq.~\eqref{eq:PM}.
The final posterior distribution given by the Bayesian theorem becomes $\tilde{\mathcal{P}}(f_c) = \mathcal{N}  \tilde{\mathcal{L}}(f_c) P(f_c)$.
Thus, the estimation of $f_c$ is given as the expectation and uncertainty within the interval $[f_l,f_r]$
\begin{align}
    \tilde{f}_{\text{est}}(\{\mu_k\}) &= \int_{f_l}^{f_r}{ \tilde{\mathcal{P}}(f_c) f_c  df_c}, \\
    \Delta^2\tilde{f}_{\text{est}}(\{\mu_k\}) &= \int_{f_l}^{f_r}{ \tilde{\mathcal{P}}(f_c) [f_c- \tilde{f}_{\text{est}}(\{\mu_k\})]^2 df_c},
\end{align}
and the corresponding mean-square error becomes
\begin{equation}
    \tilde{\varepsilon}^2 (f_{\text{est}}) = {\sum_{\{\mu_k\}} \tilde{\mathcal{L}}( f_c )[\tilde{f}_{\text{est}} (\{\mu_k\})-f_c ]^2}.
\end{equation}

\subsection{Credible-interval-based adaptive Bayesian quantum frequency estimation \label{sec:the CI-adaptive protocol}}
Entanglement-enhanced sensors based on GHZ states face a fundamental trade-off between sensitivity and dynamic range. 
Our protocol mitigates this limitation by adaptively controlling the Ramsey interrogation time $T_j$ based on the credible interval of the current posterior distribution, as shown in Fig.~\ref{fig:1}~(d).
Most previous studies~\cite{cappellaro2012spinbath,waldherr2012highdynamicrange,nusran2012highdynamicrange,cappellaro2012spinbath,bonato2016optimized,Craigie2021} employ a sequence of exponentially increasing interrogation times $T_j = 2^{j-1}T_{\text{min}}$, which also requires varying measurement copies (times) as the interrogation time increases.
This is easy to achieve with sequential measurements. However, for both parallel and cascading strategies, discarding some copies at each step results in a waste of resources.

Within the Bayesian framework, the next interrogation time $T_{j+1}$ is determined by matching the period of the subsequent likelihood function to the width of the credible interval of the current posterior distribution [as shown in the right panel of Fig.~\ref{fig:1}(d)]:
\begin{equation}
    \label{eq:credible}
    \frac{1}{N_0 T_{j+1}} = 2g\Delta f_{\text{est}}^{(j)}.
\end{equation}
Here, $\Delta f_{\text{est}}^{(j)}$ is the uncertainty (standard deviation) of the current estimation, $N_0$ is the number of particles in the smallest ensemble, and $g$ is a scaling factor associated with the desired credible level $\zeta$, which satisfies
\begin{equation}
    \int_{f_{l}}^{f_r} \tilde{P}(f_c) \, df_c = \zeta,
\end{equation}
where $f_{l,r} = f_{\text{est}}^{(j)} \pm g\Delta f_{\text{est}}^{(j)}$.
The factor $g$ depends on both the credible level $\zeta$ and the total number of measurement copies $M = \sum_k M_k$. For a fixed $M$, $g$ increases with $\zeta$. Conversely, for a fixed $\zeta$, $g$ decreases as $M$ increases. For example, when $\zeta = 99.9\%$, we have $g = 5.041$ for $M = 9$, whereas $g \approx 3.3$ when $M \to \infty$ (see Supplementary Information for more details). The scaling factor $\alpha$ can subsequently be derived from $g$ and the experimental parameters via the relation $\alpha = \frac{\kappa \pi}{g} \frac{\sqrt{N_t}}{N_0}$.

Combining this update rule with the Cram\'{e}r-Rao lower bound (CRLB) under adaptive estimation $\Delta f_{\text{est}}^{(j)} \approx \Delta f_{\text{CRLB}}^{(j)} = 1/(2\pi \kappa \sqrt{N_t} \sqrt{\sum_{i=0}^j T_i^2})$ (where $\kappa = \sqrt{\sum_k M_k N_k^2 / N_t}$ and $N_t = \sum_k M_k N_k$), we derive the optimal interrogation time sequence (see Supplementary Information for more details):
\begin{align}
     T_j=  T_0 \alpha \left(\sqrt{1+\alpha^2}\right)^{j-1}, j\ge 1,
\end{align}
with the scaling factor $\alpha (\zeta) = \frac{\kappa \pi}{g(\zeta)} \frac{\sqrt{N_t}}{N_0}$.
Starting from the minimum interrogation time $T_0 = T_{\text{min}}$, and subject to the constraint $T_{\text{min}} \leq T_j \leq T_{\text{max}}$, the predetermined interrogation time is derived as follows:
\begin{equation}
\label{eq:Tj_sequenc}
T_j = \min \left\{ \max \left[ T_{\text{min}},\; T_{\text{min}} \alpha \left(\sqrt{1+\alpha^2}\right)^{j-1}\right],\; T_{\text{max}} \right\}.
\end{equation}
The selection of $T_{\text{min}}$ and $T_{\text{max}}$ involves a trade-off between the desired dynamic range amplification factor $T_{\text{max}}/T_{\text{min}}$ and the experimental capabilities. 
The value of $T_{\text{min}}$ is fundamentally limited by the time resolution of the instrument, whereas $T_{\text{max}}$ is constrained by the coherence time of the system. The effect of dephasing noise on the maximum interrogation time will be discussed in Sec.~\ref{sec:dephasing}.

This sequence yields the following theoretical uncertainty scaling:
\begin{subnumcases}{\label{eq:delta_ft} \Delta f_{\text{CRLB}} =}
    \label{eq:delta_f1}
    \dfrac{1}{2\pi \kappa \sqrt{N_t}} \dfrac{1}{ T_{\text{min}} + (t - T_{\text{min}}) \beta } & \(T_j \le T_{\text{max}}\), \\
    \label{eq:delta_f2}
    \dfrac{1}{2\pi \kappa \sqrt{N_t}} \dfrac{1}{\sqrt{t T_{\text{max}}}} & \(T_j = T_{\text{max}}\),
\end{subnumcases}
where $t = \sum_{i=0}^j T_i$ is the total interrogation time and $\beta= \sqrt{1+\alpha^{-2}} - \alpha^{-1}$. 

\textit{Dual Heisenberg scaling:} 
For $\alpha = 1$ and $T_{\text{min}} < T_j < T_{\text{max}}$ and $t \gg T_{\text{min}}$, Eq. \eqref{eq:delta_f1} can be simplified to:
\begin{equation}
\label{eq:dual-HL}
\Delta f_{\text{est}} \approx \dfrac{(\sqrt{2} + 1)}{2\pi \kappa \sqrt{N_t} t}.
\end{equation}
For fixed $M$ copies of an $N$-qubit GHZ state ($\kappa\sqrt{N_t} = \sqrt{M}N$), this achieves \textit{dual Heisenberg scaling} $\Delta f_{\text{est}} \propto 1/(N t)$ with respect to both particle number $N$ and total interrogation time $t$. 
Once $T_j$ saturates in $T_{\text{max}}$, the scaling becomes hybrid: Heisenberg scaling versus $N$ but SQL versus $t$, i.e., $\Delta f_{\text{est}} \propto 1/(N \sqrt{t})$.
The protocol with $\zeta>99.999\%$ and $M \geq 9$ copies per step enables robust performance within the whole dynamic range (see Supplementary Information for more details).

The general implementation of CI-adaptive can be seen in the following Algorithm \ref{alg:CI-adaptive}, and its implementation with cascaded GHZ states (including identical-size and different-size) is detailed in the next section.\\

\begin{algorithm}[H]
\SetKwInOut{Initialize}{Initialize}
\caption{CI-adaptive protocol overview}\label{alg:CI-adaptive}
\Require{minimum interrogation time $T_{\text{min}}$; maximum interrogation time $T_{\text{max}}$; iteration steps $n$; CI-factor $\alpha$; Ensembles: $\{N_{k=0}^{K},M_{k=0}^{K}\}$ \;}
\Initialize{Prior $P^{(0)}(f_c)$, LO frequency $f^{(0)}_L$ }
\For{$j = 0$ \textbf{to} $n$}{
    $T_j=\min \left\{ \max \left[ T_{\text{min}}, T_{\text{min}} \alpha \left( \sqrt{1+\alpha^2} \right)^{j-1} \right], T_{\text{max}} \right\}$\;
    \For{$k=0$ \textbf{to} $K$}{
        $\theta_{0}=\pi/2,\theta_{k>0}=0$\;
        $\mu_{k} = \text{Ramsey}(N_k,M_k,\theta_{k},T_j,f_L^{(j)})$\;
        
        $\text{Likelihood update}\; \mathcal{L}^{(j)}_{k}(N_k, M_k,\mu_{k},\theta_{k},T_j,f_L^{(j)}|f_c)$ 
    }
    $\text{Posterior update}\; \mathcal{P}^{(j)}(f_c|\{N_k\}, \{M_k\},\{\mu_{k}\},\{\theta_{k}\},T_j,f_L^{(j)})$\;
    $\text{Estimation} \; f_{\text{est}}^{(j)} = \int_{f_l}^{f_r}{ \mathcal{P}^{(j)}(f_c|\cdots) f_c df_c}$\;
    LO frequency update $f_L^{(j+1)}\leftarrow f_{\text{est}}^{(j)}$\;
    Prior update $P^{(j+1)}(f_c) \leftarrow  \mathcal{P}^{(j)}(f_c|\cdots) $\;
}
\Ensure{Posterior distribution $\mathcal{P}^{(n)}(f_c|\cdots) $; Estimation $f_{\text{est}}^{(n)}$ with variance $\Delta f_{\text{est}}^{(n)}$\;}
\end{algorithm}

\section{Results \label{sec:Results}}

In this section, based on the designed time update rules, we give concrete implementation examples for entanglement-enhanced atomic clocks, including individual GHZ states and cascaded GHZ states.
Generally, the likelihood function and the posterior distribution for the $K+1$ ensembles are given respectively as $\tilde{\mathcal{L}}(f_c)=\prod_{k=0}^{K}{\mathcal{L}_{M_k}(\mu_k|N_k,\theta_k,f,f_L,T)}$ and $\tilde{\mathcal{P}}(f_c) = \tilde{\mathcal{N}} \tilde{\mathcal{L}}(f_c) P(f_c) $, where $k=0,1,...,K$ represents different ensembles and $\tilde{\mathcal{N}}$ is the normalization factor. 
Since it is difficult to go through all possible outcomes within the framework of the CI-adaptive protocol, hereafter we use the average of multiple simulations for evaluation.
Therefore, the estimated frequency is given as $\bar{f}_{\text{est}}^{(j)} = \frac{1}{R}\sum_{r=1}^{R}{[f_{\text{est}}^{(j)}]_{r}}$, where $R$ is the repetition times.
Then the related average uncertainty is also replaced by $\Delta\bar{f}_{\text{est}}^{(j)} = \frac{1}{R}\sqrt{\sum_{r=1}^{R}{\{[f_{\text{est}}^{(j)}]_r}- \bar{f}_{\text{est}}^{(j)}\}^2 }$ and $\bar{\varepsilon}({f}_{\text{est}}^{(j)}) = \frac{1}{R}\sqrt{\sum_{r=1}^{R}{\{[f_{\text{est}}^{(j)}}]_r-f_c\}^2 }$.
In our simulation, $R$ is chosen as $R=5000$.

\begin{figure*}[tbh]
    \centering    \includegraphics[width=\linewidth]{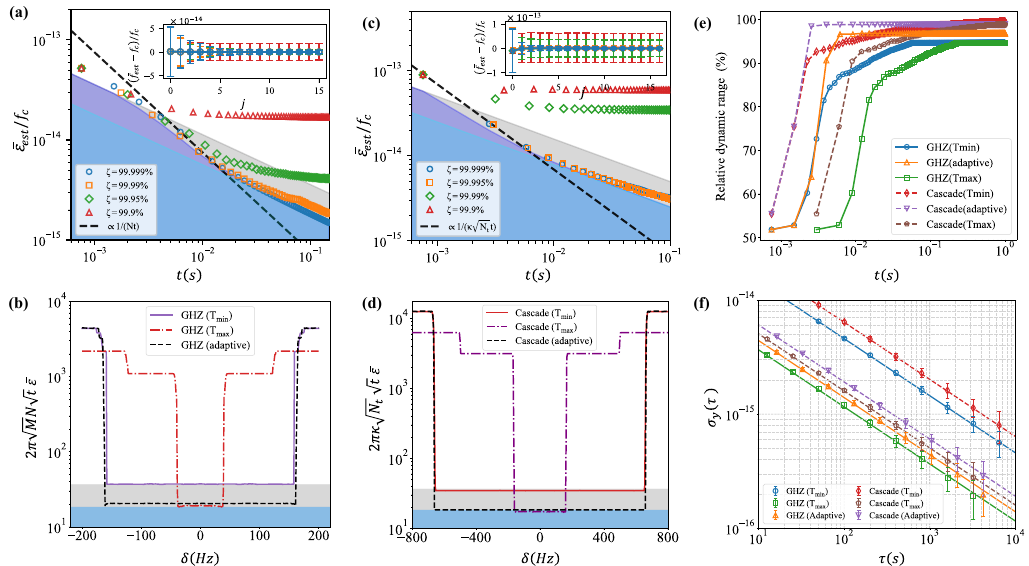}
    \caption{ 
    \textbf{Performances of our credible-interval-based adaptive Bayesian protocol for entanglement-enhanced atomic clocks.}  
    (a) Root-mean-square error (RMSE) versus total interrogation time for cascading of two groups of GHZ states under different credible level $\zeta$.
    The corresponding $\alpha(\zeta)$ are $\alpha(99.999\%)=0.963$, $\alpha(99.99\%)=1.3237$, $\alpha(99.95\%)=1.6778$, and $\alpha(99.9\%)=1.86951$.
    The particle number and copy number are $(N_0=4,M_0=4)$ and $(N_1 = 4, M_1 = 5)$, respectively.
    The three colored regions indicate the Cram\'{e}r-Rao lower bounds (CRLB) for: 
    the frequentist scheme with fixed $T_{\text{min}}=0.75$ ms (gray, top region),
    the frequentist scheme with fixed $T_{\text{max}}=3$ ms (blue, bottom region),
    and the adaptive scheme for $T_{\text{min}} \rightarrow T_{\text{max}}$ (purple).
    The intermediate position of the purple region illustrates the transition from SQL to Heisenberg-limited scaling.
    The black dashed line corresponds to the theoretical dual Heisenberg scaling (Eq. 10). 
    Inset: Bias versus iteration steps with different credible level $\zeta$. 
    (b) RMSE versus initial detuning $\delta$ for frequentist schemes (fixed $T_{\text{min}}$: $40$ steps; fixed $T_{\text{max}}$: $10$ steps) and CI-adaptive scheme in (a) ($\zeta=99.999\%$, $13$ steps). Dashed line is the theoretical lower bound. 
    (c) RMSE versus total interrogation time for cascaded GHZ states ($N_k = \{1,1,2,4\}$, $M_k = \{7,7,7,2\}$) under different credible level $\zeta$.  The corresponding $\alpha(\zeta)$ are $\alpha(99.999\%)=2.64611$, $\alpha(99.995\%)=3.00035$, $\alpha(99.99\%)=3.1812$, and $\alpha(99.9\%)=3.97311$. 
    Hybrid scaling (Heisenberg in $N$, SQL in $t$) emerges when $T_j$ saturates at $T_{\text{max}}$ (blue region), while adaptive protocol (purple) maintains dual scaling until saturation.
    (d) RMSE versus initial detuning $\delta$ for frequentist schemes (fixed $T_{\text{min}}$: $40$ steps; fixed $T_{\text{max}}$: $10$ steps) and CI-adaptive scheme in (c) ($\zeta=99.999\%$, $11$ steps). 
    (e) Relative dynamic range (frequency range where RMSE $\leq$ 1.1$\times$CRLB) versus total interrogation time. 
    (f) Overlapping Allan deviation $\sigma_y(\tau)$ versus averaging time $\tau$ showing clock stability. 
    Error bars indicate $\pm$1 standard deviation from $1000$ simulations. Data in (a-e) are averaged with $R=5000$ simulations. Data in (f) are averaged with $R=1000$ simulations.
    The $^{88}$Sr clock transition $f_c \approx 4.295\times10^{14}$ Hz.
    }
    \label{fig:2}
\end{figure*}

\subsection{Cascaded identical and different-size GHZ states}

The CI-adaptive protocol, schematically illustrated in Fig.~\ref{fig:1}(d), is universally applicable to both identical-size and different-size GHZ state configurations. 
For the cascading of identical-size GHZ states ($K=1$), we utilize two $N$-particle ensembles with $M_0 = \lfloor M/2 \rfloor$ copies using an auxiliary phase of $\theta_0 = \pi/2$ and $M_1 = M - M_0$ copies using $\theta_1 = 0$. 
For the cascading of different-size GHZ states, the number of particles grows exponentially as $N_k = 2^{k-1}N_0$, while the number of copies decreases linearly as $M_k = M_K + v(K-k)$~\cite{higgins2007entanglementfree,berry2009how}. 
In this configuration, the auxiliary phase is applied exclusively to the first smallest ensemble ($N_0$). For the smallest ensembles ($N_1 = N_0$), the copy numbers are given by $M_0 = \lfloor \frac{M_K + v(K - 1)}{2} \rfloor$ and $M_1 = M_K + v(K - 1) - M_0$, following the same assignment rule as in the identical-size case.

As shown in Figs.~\ref{fig:2}(a-d), we present concrete implementations of entanglement-enhanced atomic clocks based on the CI-adaptive method. 
These include cascading with identical GHZ states ($N_0 = N_1 = 4$, $M_0 = 4$, $M_1 = 5$) and with different-size GHZ states ($N_k = \{1,1,2,4\}$, $M_k = \{7,7,7,2\}$).
A uniform prior distribution $P(f_c) = 1/T$ is sufficient for initialization, since the auxiliary phase is incorporated. This choice is particularly appropriate when the phase to be estimated is entirely unknown. 
As outlined in Fig.~\ref{fig:1}(d) and Algorithm~\ref{alg:CI-adaptive}, three key parameters are adaptively updated during the estimation process: the local oscillator frequency $f_L^{(j)}$ is set to the previous estimate $f_{\text{est}}^{(j-1)}$, the interrogation time $T_j$ is determined by Eq.~\eqref{eq:Tj_sequenc}, and the prior distribution $P^{(j)}(f_c)$ is updated using the posterior distribution $\tilde{\mathcal{P}}(f_c)$ from the previous step.

The results of the above implementations demonstrate three key advances:
\begin{enumerate}[i)]
    \item \textit{Dual Heisenberg scaling}: When the interrogation time is between the minimum and maximum bounds ($T_{\text{min}} < T_j < T_{\text{max}}$), the precision scales as $\Delta f_{\text{est}} \propto 1/(Nt)$ with respect to both the particle number $N$ and the total interrogation time $t$, as shown in Figs.~\ref{fig:2}(a) and (c). 
    The purple region in Fig.~\ref{fig:2}(a) represents the Cram\'{e}r-Rao lower bound (CRLB) for the adaptive protocol, which lies between the CRLBs for the fixed-$T_{\text{min}}$ (gray region) and fixed-$T_{\text{max}}$ (blue region) cases, illustrating the transition from the SQL scaling to the Heisenberg scaling. 
    The theoretical dual Heisenberg scaling derived from Eq.~\eqref{eq:dual-HL} is indicated by the black dashed line.

    \item \textit{Hybrid scaling}: When $T_j$ reaches the maximum interrogation time $T_{\text{max}}$, the scaling law is changed to $\Delta f_{\text{est}} \propto 1/(N\sqrt{t})$. This represents a hybrid scaling: Heisenberg scaling with respect to $N$ but SQL with respect to $t$, as evidenced in Figs.~\ref{fig:2}(a) and (c). 
    The corresponding CRLB for this hybrid scaling regime is represented by the blue region in Fig.~\ref{fig:2}(c).
    
    \item \textit{Dynamic range expansion}: Our protocol achieves a $(T_{\text{max}}/T_{\text{min}})$-fold extension of the dynamic range while maintaining a precision comparable to that obtained using $T_{\text{max}}$ alone in entanglement-enhanced atomic clocks, see Figs.~\ref{fig:2}(b), (d), and (e). 
    By employing cascaded different-size GHZ states, the dynamic range can be further extended by a factor of $(\frac{N_{\text{max}} T_{\text{max}}}{N_{\text{min}} T_{\text{min}}})$. The relative dynamic range — defined as the frequency range over which the root-mean-square error (RMSE) remains within $1.1$ times the CRLB — increases with the total interrogation time, as demonstrated in Fig.~\ref{fig:2}(e).
\end{enumerate}

In designing the time sequence, the choice of credible level $\zeta$ governs the growth rate of the interrogation time. 
A higher $\zeta$ requires a longer interrogation time to achieve a wider (more certain) interval.
While a lower $\zeta$ accelerates convergence to the maximum interrogation time, it has a higher risk of period-skipping errors at the expense (see Supplementary Information for more details).
To prevent period-skipping errors, the credible level should be high enough, e.g., $\zeta \geq 99.999\%$, which corresponds to $\alpha = 1$ and $\alpha = 2.70154$ for the individual and cascaded GHZ states in Figs.~\ref{fig:2}~(a) and (c). 
While our protocol employs a predetermined time sequence based on a fixed credible width, future extensions could incorporate real-time updates of credible intervals for enhanced error resilience.

\subsection{Relative dynamic range}
The CI-adaptive protocol effectively extends the dynamic range to the period length determined by the minimum particle number and the minimum interrogation time. 
As illustrated in Figs.~\ref{fig:2}~(b) and (d), the RMSE approaches the CRLB across nearly the entire phase range, exhibiting significant deviations only at the boundaries. 
These deviations originate from the property of the mean-square error: its first term is governed by the estimation bias, while the expected value of the posterior distribution in the Bayesian estimator fails to properly manifest near the boundaries.
Consequently, there will always be a substantial bias at the boundaries, leading to an increased RMSE.

To effectively observe changes in the dynamic range, we define the relative dynamic range as $\frac{\Delta f_{dyn}}{1/(NT_{\text{min}})}$, where $\Delta f_{dyn}$ represents the achievable dynamic range.
In the achievable dynamic range, the mean-square error $\bar{\varepsilon}^2_{\text{est}}$ does not exceed 10\% above the Cram\'{e}r-Rao bound: $\frac{\bar{\varepsilon}^2_{\text{est}} -\Delta^2 f_{CRLB}}{{\varepsilon}^2_{\text{est}}} \le 10\% $.
For different configurations, the relative dynamic range takes different values.
For individual GHZ states with fixed interrogation time $T_{\text{max}}$, the relative dynamic range is $\frac{\Delta f_{dyn}}{1/(NT_{\text{max}})}$.
For individual GHZ states with fixed interrogation time $T_{\text{min}}$ or growing from $T_{\text{min}}$ according to the CI-adaptive protocol, the relative dynamic range is $\frac{\Delta f_{dyn}}{1/(NT_{\text{min}})}$.
For cascaded GHZ states with fixed interrogation time $T_{\text{max}}$, the relative dynamic range is $\frac{\Delta f_{dyn}}{1/T_{\text{max}}}$.
For cascaded GHZ states with fixed interrogation time $T_{\text{min}}$ or growing from $T_{\text{min}}$ according to the CI-adaptive protocol, the relative dynamic range is $\frac{\Delta f_{dyn}}{1/T_{\text{min}}}$.
In Figs.~\ref{fig:2}~(c) and (d), we show the dynamic range for different configurations. 
It features a near-horizontal central segment during the interval, followed by sharp roll-offs at both edges.
It clearly shows that the adaptive protocol extends the dynamic range as the total interrogation time increases, see Fig.~\ref{fig:2}~(e).
Notably, the CI-adaptive protocol achieves the dynamic range threshold with substantially reduced interrogation time compared to conventional methods, for both individual and cascaded GHZ states.

\subsection{Stability analysis}
The clock frequency is locked based on the output of a CI-adaptive protocol cycle, which consists of $n$ steps of adaptive measurements.
To evaluate the sensitivity, we calculate the fractional frequency stability $y = f_{\text{est}}/f_c$ after different steps of the CI-adaptive protocol cycle.
In each locking cycle, the interrogation time and the prior distribution are reset, while the LO frequency is updated to $f_{\text{est}}^{(n)}$ obtained from the previous cycle.
Assuming that the measurements of the $N_k$ ensembles in the CI-adaptive protocol are implemented simultaneously with $M_k$ copies at each step, the duration of each locking cycle is given by
\begin{equation}
    T_{\text{cycle}} = \sum_{j=0}^{n} (T_j + T_D),
\end{equation}
where $T_D$ is the dead time for each step and $t = \sum_{j=0}^{n} T_j$ is the total interrogation time that determines the uncertainty of the output estimation in one cycle according to Eq.~\eqref{eq:delta_ft}.

The overlapping Allan deviation of the fractional frequency $y = f_{\text{est}}/f_c$ is theoretically given by~\cite{riehle2003frequency}
\begin{equation}
    \label{eq:allanvar_CAS}
    \sigma_y(\tau) = \frac{1}{2\pi \kappa \sqrt{N_{t}} f_c \sqrt{t T_{\text{max}}}} \sqrt{\frac{T_{\text{cycle}}}{\tau}},
\end{equation}
which characterizes the clock stability. 
Here, $\tau$ represents the total averaging time of the locking process, which spans multiple cycles.

For comparison, we give $\sigma_y(\tau)$ for both frequentist and Bayesian schemes employing either identical- or different-size GHZ states.
In our calculations, we select $n=40$ for $T = T_{\text{min}}$ and $n=10$ for $T = T_{\text{max}}$.
This choice ensures that the total interrogation time $t = 30$ ms closely matches that of the CI-adaptive protocol scheme with $n=13$ and $\alpha=1$, where $t \approx 30.18$ ms.
According to Ref.~\cite{cao2024multiqubit}, we set $T_{\text{max}} + T_D = 1.26$ s with a dead time $T_D = 1.257$ s.
Averaged over 1000 simulations, the CI-adaptive protocol scheme using same-size GHZ states achieves a stability of $1.3(9) \times 10^{-14}/\sqrt{\tau}$. This represents a $10.3(2)$ dB improvement over the conventional scheme with $T = T_{\text{min}}$ and is only $1.6(3)$ dB worse than the conventional scheme with $T = T_{\text{max}}$, as shown in Fig.~\ref{fig:2}(f).
The CI-adaptive protocol scheme with cascaded GHZ states yields a stability of $1.9(1) \times 10^{-14}/\sqrt{\tau}$, which is slightly inferior to the case of cascaded GHZ states using $T = T_{\text{max}}$ but significantly better than the case using $T = T_{\text{min}}$.
As $n$ increases, the stability approaches that of frequentist measurements with $T = T_{\text{max}}$, albeit at the cost of increasing $T_{\text{cycle}}$.

\subsection{Influence of dephasing\label{sec:dephasing}}

Dephasing-induced precision impairment represents a major challenge for quantum metrology using GHZ states. Here, we analyze the effect of individual dephasing during the free interrogation process in Ramsey interferometry, governed by the Hamiltonian $H_0 = \delta \hat{J}_z$.
In the presence of dephasing, the system's time evolution is described by the master equation~\cite{huelga1997improvement,kielinski2024ghz},
\begin{equation}
    \label{eq:master}
    \dot{\rho} = -i[\hat{H}_0, \rho] + \frac{\gamma}{2} \sum_{k=1}^N \mathcal{L}_{\hat{J}_z^{(k)}}[\rho],
\end{equation}
where the Lindblad superoperator is defined as $\hat{L}_{\hat{A}}[\rho]= 2\hat{A}\rho \hat{A}^{\dagger} - \rho\hat{A}^{\dagger}\hat{A} - \hat{A}^{\dagger}\hat{A}\rho $, with $\hat{J}_z^{(k)}=\sigma_z^{(k)}/2$ and single-particle Pauli operator $\sigma_z^{(k)}$.
Here, $\rho$ denotes the density operator and $\gamma$ represents the individual dephasing rate.

Although the accumulated phase may be amplified by a factor of $N$, the decoherence time of the GHZ state is reduced to $1/N$ of that for spin coherent states.
Consequently, the measurement probability becomes
\begin{equation}
    \label{eq:dephasing_prob}
    \mathcal{L}_u = \frac{1}{2} \left\{ 1 + e^{-\frac{\gamma}{2} N T} \times \sin[2\pi N T (f_c - f_L) + \theta] \right\}.
\end{equation}
The coherence time for a single-particle system ($N=1$) is defined as $T_2^* = 2 / \gamma$. 
For an $N$-particle GHZ state, the effective coherence time reduces to $T_{\text{dec}} = T_2^* / N$ due to collective decoherence effects. 
Thus dephasing becomes increasingly dominant as $N$ grows, severely limiting the maximum achievable interrogation time.

Considering $M$ copies of identical GHZ states, the frequency estimation uncertainty under a maximum interrogation time $T_{\text{max}}$ constraint is given by (see Supplementary Information for the derivation):
\begin{equation}
    \Delta f_{\text{CRLB}} = \frac{1}{2\pi \sqrt{M} N} \cdot \frac{1}{\sqrt{\sum_{i=0}^{j}{T^2 e^{-\gamma N T}} }} \ge \frac{1}{2\pi \sqrt{M} N} \cdot \frac{e^{\gamma N T_{\text{max}}/2}}{\sqrt{t T_{\text{max}}  }} .
    \label{eq:uncertainty_dephasing}
\end{equation}
Minimizing this uncertainty with respect to the interrogation time $T$ yields the optimal maximum interrogation time~\cite{huelga1997improvement,kielinski2024ghz},
\begin{equation}\label{Topt}
    T_{\text{max}}^{\text{opt}} = \frac{1}{\gamma N} = \frac{T_2^*}{2N}=\frac{T_{dec}}{2}.
\end{equation}

Thus, the maximum interrogation time should satisfy $T_{\text{max}} \lesssim T_{\text{max}}^{\text{opt}}=T_{dec}/2$.
Interrogation durations beyond $T_{\text{max}}^{\text{opt}}$ place the system deep into the decoherence-dominated regime, where signal quality (encoded in the exponential factor $e^{\gamma N T/2}$) deteriorates faster than SQL scaling ($1/\sqrt{T}$) can compensate, leading to increased estimation uncertainty.

\begin{figure}[H]
    \centering    
    \includegraphics[width=\columnwidth]{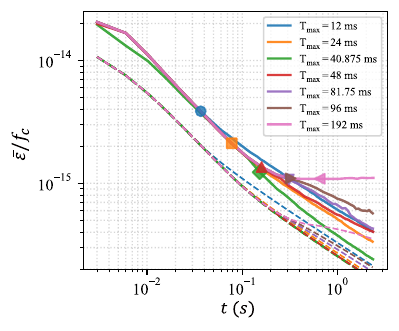}
    \caption{ 
    {\bf Influence of dephasing on the fractional uncertainty.} 
    Simulations are performed with the CI-adaptive protocol under different maximum interrogation time $T_{\text{max}}$ for $M=9$ copies of GHZ states ($N=4$) with $T_{\text{min}}=3$ ms, $\alpha=1$, and $R=1024$. 
    The dashed lines represent the theoretical CRLB as Eq.~\eqref{eq:uncertainty_dephasing}, the solid lines represent simulation RMSE during the adaptive process, and the marks ({\fontsize{15pt}{14pt}${\circ}$}$: T_{\text{max}}=12$ ms, $\square: T_{\text{max}}=24$ ms, $\Diamond: T_{\text{max}}=40.875$ ms, $\triangle: T_{\text{max}}=48$ ms, $\triangledown: T_{\text{max}}=81.75$ ms, $\rhd: T_{\text{max}}=96$ ms, $\lhd: T_{\text{max}}=192$ ms) indicate the positions where the corresponding maximum interrogation time is reached.}
    \label{fig:3}
\end{figure}

Considering the experimental situation, we take the 4-qubit-GHZ state as an example.  
Its coherence time ($T_{dec}=T_2^*/4=81.75$ ms) is $1/4$ of the coherence time of uncorrelated particles ($T_2^* = 327$ ms)~\cite{cao2024multiqubit}.
In previous sections, we have demonstrated the adaptive Bayesian estimation with the interrogation time growing from $0.75$ ms to $3$ ms, which makes the GHZ state have the same dynamic range as the spin coherent state while maintaining the high precision.
The interrogation time of less than $3$ ms ensures that the results are nearly unaffected by noise.
However, $3$ ms is a relatively short Ramsey interrogation time and is not the longest interrogation time achievable with the 4-qubit GHZ state in experiments.
Here, taking into account the coherence time, we further demonstrate that adaptively increasing the interrogation time from $3$ ms to $T_{\text{max}}$ (where $T_{\text{max}}$ can exceed the coherence time) allows us to explore how to achieve the best precision within our protocol.

As shown in Fig.~\ref{fig:3}, we present the fractional uncertainty versus the interrogation time $t$ for different maximum interrogation times.
All results have similar trends before reaching the corresponding $T_{\text{max}}$ or $T_{dec}/2$.
If $T_{\text{max}} $ is rather smaller than the coherence time, the precision will decrease to the SQL scaling after reaching $T_{\text{max}}$. 
However, if $T_{\text{max}}$ is chosen close to the coherence time, it will be the scaling worse than SQL after reaching $T_{\text{max}}$. 
When $T_{\text{max}}$ is much greater than the coherence time, increasing the measurement time cannot continue to improve the precision and will remain at the level when it reaches the coherence time.
From our numerical results, the optimal performance in practical implementations requires setting the maximum interrogation time, $T_{\text{max}}\lesssim T_{dec}/2$, which is consistent with the theoretical condition of Eq.~\eqref{Topt}.

\subsection{Comparisons with conventional Fourier-coefficients-based protocols \label{sec:protocols}}
\begin{figure*}
    \centering  \includegraphics[width=1.0\linewidth]{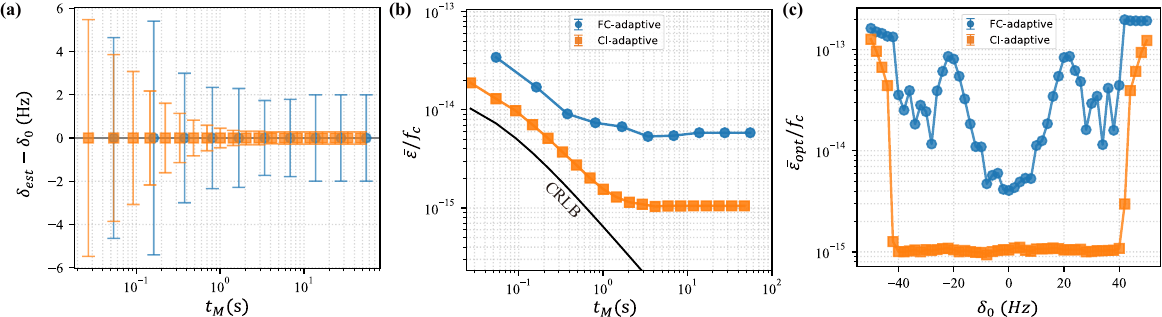} \caption{{\bf Comparisons between the CI-adaptive protocol and the FC-adaptive protocol}. The CI-adaptive protocol ($M=9$ copies, $n=20$ steps) and the FC-adaptive protocol ($M=18$ copies, $n=10$ steps) are simulated with $R=1024$ times under the same total time resource $t_M=M\sum_{j=0}^n T_j$. Here, $T_{\text{min}}=3$ ms, $T_{\text{max}}=2^n T_{\text{min}}$. (a) The mean error of estimation versus the total time resource. (b) The variation of the averaged mean-square error in logarithmic scale versus the total time resource. (c) The optimal averaged mean-square error versus the detuning $\delta$ within the dynamic range.}
    \label{fig:4}
\end{figure*}

Here, we compare our protocol with conventional ones to show its exceptional advances. 
Conventional protocols employ exponentially increasing interrogation times with a growth rate of $2$ (i.e. $T_j = 2^{j} T_{\text{min}}$) in adaptive Bayesian estimation~\cite{nusran2012highdynamicrange,waldherr2012highdynamicrange} to optimize phase estimation by minimizing Holevo variance~\cite{holevo1984covarianta}.
The exponent rate $2$ enables one to update the Fourier coefficients of the probability density~\cite{said_Nanoscale_2011}.  
Subsequent studies have shown that increasing the number of measurement repetitions at shorter interrogation times can further enhance performance~\cite{cappellaro2012spinbath,bonato2016optimized}.
This Fourier-coefficients-based method exhibits a relatively rapid growth rate in time and can achieve good performance when the copy number scales linearly with each step.
However, for cascaded GHZ states, experimental constraints necessitate using a fixed number of identical state copies per interrogation step.
While the Holevo variance is a valuable metric for periodic distributions, we adopt the mean-square error in our analysis because it provides a more comprehensive measure.
The mean-square error incorporates both variance and bias components, providing a comprehensive characterization of estimation performance by simultaneously quantifying precision and accuracy.

In the context of fixed copies, we perform a comparative analysis between our CI-adaptive protocol and the FC-adaptive protocol~\cite{bonato2016optimized,Craigie2021}, see Algorithm~\ref{alg:FC-adaptive}.\\

\begin{algorithm}[H]
\SetKwInOut{Initialize}{Initialize}
\caption{FC-adaptive protocol overview}\label{alg:FC-adaptive}
\Require{minimum interrogation time $T_{\text{min}}$; maximum interrogation time $T_{\text{max}}$; iteration steps $n$;  Ensembles: $\{N_{k=0}^{K},M_{k=0}^{K}\}$ \;}
\Initialize{Prior $P^{(0)}(f_c)$, LO frequency $f_L$ }
\For{$j = 0$ \textbf{to} $n$}{
    $T_j = \min[2^j T_{\text{min}}, T_{\text{max}}]$\\ 
    \textit{Choose $\theta_j=\frac{1}{2} \arg{\{p_{2^j}\}}$}\\ 
    \For{$k=0$ \textbf{to} $K$}
    {
        $\mu_{k} = \text{Ramsey}(N_k,M_k,\theta^{(j)},T_j,f_L)$\;
        
        $\text{Likelihood update}\; \mathcal{L}^{(j)}_{k}(N_k, M_k,\mu_{k},\theta^{(j)},T_j,f_L|f_c)$ 
    }
    $\text{Posterior update}\; \mathcal{P}^{(j)}(f_c|\{N_k\}, \{M_k\},\{\mu_{k}\},\{\theta_{k}\},T_j,f_L)$\;
    $\text{Estimation} \; f_{\text{est}}^{(j)} = \int_{f_l}^{f_r}{ \mathcal{P}^{(j)}(f_c|\cdots) f_c df_c}$\;
    Prior update $P^{(j+1)}(f_c) \leftarrow  \mathcal{P}^{(j)}(f_c|\cdots) $\;
}
\Ensure{Posterior distribution $\mathcal{P}^{(n)}(f_c|\cdots) $; Estimation $f_{\text{est}}^{(n)}$ with variance $\Delta f_{\text{est}}^{(n)}$\;}
\end{algorithm}

In the FC-adaptive protocol, the auxiliary phase is updated by the coefficient $p_{\lambda=2^j}$ of the prior probability distribution in the Fourier space,
\begin{equation}
    \label{eq:p_k}
    P^{(j)}(\phi) = \sum_{\lambda}{p_{\lambda}^{(j)}e^{i\lambda\phi}}.
\end{equation}
Here, we use the 4-qubit-GHZ states for comparison and take the factor $\alpha=1$. 
Consequently, the CI-adaptive protocol doubles the number of steps to reach the same interrogation time as the FC-adaptive protocol. 
To maintain the same total resource utilization (with total copies $M_t=M\times n$), we set the copy number for the FC-adaptive protocol to be twice that of the CI-adaptive protocol. 
The total time resource is defined correspondingly as $t_M=\sum_j{M T_j}$.

In Fig.~\ref{fig:4}, we compare the CI-adaptive protocol of $M=9$ copies and $n=20$ steps with the FC-adaptive protocol of $M=18$ copies and $n=10$ steps.
In our analysis, we incorporate the dephasing effects according to Eq.~\eqref{eq:dephasing_prob}, with $T_{\text{min}}=3$ ms and $T_{\text{max}}=2^n T_{\text{min}}$.
As established in Sec.~\ref{sec:dephasing}, extending the interrogation time beyond the system's coherence time yields no further metrological benefit, since decoherence fundamentally constrains the extractable information.
Given identical total copy resources, the CI-adaptive protocol achieves superior precision than the FC-adaptive protocol with relatively fewer copies but more steps, which shows its advantages in the case of few copies and in the presence of dephasing noise.
Beyond its precision advantages, our CI-adaptive protocol uniquely incorporates the posterior distribution at each step for real-time adaptive feedback.
Unlike the FC-adaptive protocol, which requires Fourier space conversion to calculate coefficients, our approach reduces computational overhead and achieves faster system response in practice.

\section{Conclusion and discussion\label{sec:conclusion}}
In summary, our work presents an adaptive Bayesian quantum frequency estimation protocol based on credible intervals. 
This protocol enables GHZ-state atomic clocks to combine a high dynamic range with dual Heisenberg-limited precision scaling \textemdash  a critical advancement for quantum metrology.
In parallel with the use of cascaded GHZ states, our approach combines optimized Ramsey interferometry sequences with variable interrogation times, providing an effective solution to extend the dynamic range of GHZ-state sensors while maintaining quantum-enhanced performance.
In particular, the CI-adaptive protocol scheme can achieve dual Heisenberg-limited precision scaling with respect to both particle number and total interrogation time before reaching the coherence time or maximum interrogation time.  
Moreover, we show that by combining our CI-adaptive protocol with cascaded GHZ states, the dynamic range can be further extended, which can effectively overcome the trade-off between sensitivity and dynamic range in conventional GHZ-state-based quantum metrology. 

Moreover, the CI-adaptive protocol simultaneously improves both precision and dynamic range in cascaded systems while maintaining a fixed number of measurement copies (i.e., repeated measurements at constant interrogation time). 
Unlike the FC-adaptive protocol, which requires computationally intensive transformations for calculating Fourier coefficients, this approach leverages direct posterior distributions, enabling faster feedback and lower computational resource demands. 
This combination of performance and efficiency offers a practical advancement for adaptive Bayesian protocols in real-world applications.

Furthermore, within the Bayesian quantum estimation framework, combining interaction-based detection with sign measurements, one may enhance the robustness of GHZ-state-based atomic clocks against decoherence and detection noise (see Supplementary Information). 
This approach offers more promising experimental feasibility compared to traditional parity measurements.
Taking advantage of recent advances in multiqubit entanglement~\cite{cao2024multiqubit}, universal quantum control~\cite{finkelstein2024universal}, and atomic clock stability~\cite{han2024atomic}, this work establishes a practical framework for building next-generation atomic clocks that simultaneously achieve unprecedented precision and large dynamic range.
While our protocol employs a predetermined interferometry sequence derived from our theoretical framework, it also accommodates real-time adjustments of interrogation times in practical applications~\cite{leroux_Online_2017,joas_Online_2021}. 
By dynamically updating these parameters based on measured variance and credible intervals, the system may achieve enhanced noise resilience. 
Employing feedback based on variance and credible interval, this real-time adaptive approach may enable flexible operation across various sensing scenarios.
\newline

\noindent
\textbf{Supporting information}

\noindent
The supporting information can be found in the attachment at the end of this document.
\newline

\noindent
\textbf{{Conflict of interest}}

\noindent
The authors declare that they have no conflict of interest.
\newline

\noindent
\textbf{{Acknowledgments}}
Jungeng Zhou and Jiahao Huang contribute equally. The authors thank Dr. Raphael Kaubruegger very much for his helpful comments and suggestion. This work is supported by the National Natural Science Foundation of China (12025509, 12104521, 12475029, 92476201), the National Key Research and Development Program of China (2022YFA1404104), and the Guangdong Provincial Quantum Science Strategic Initiative (GDZX2305006, GDZX2405002).

%


\clearpage
\newpage

\appendices
  
\renewcommand{\thetable}{S\arabic{table}}
\renewcommand{\thefigure}{S\arabic{figure}}
\renewcommand{\theequation}{S\arabic{equation}}
\renewcommand{\thesection}{\Alph{section}}
\setcounter{table}{0}
\setcounter{figure}{0}
\setcounter{equation}{0}	
\setcounter{section}{0}	

\setcounter{algocf}{0}
\renewcommand{\thealgocf}{S.\arabic{algocf}}
\renewcommand*{\theHalgocf}{\thealgocf}



\begin{widetext}

\noindent
\begin{center}
\Large\textbf{{Supplementary Information for ``Credible-interval-based adaptive Bayesian quantum
frequency estimation for entanglement-enhanced atomic clocks''}}
\end{center}

\section{Flowing chart of locking with ``CI-adaptive" protocol\label{SM:algorithm}}

In this section, we present the basic procedure for locking the atomic transition frequency with credible-interval-based adaptive Bayesian quantum frequency estimation.
As shown in Algorithm.~\ref{alg:locking}, the LO frequency is updated by the output of the ``CI-adaptive and the estimation range is updated accordingly. The initial prior distributions for the ``CI-adaptive protocol are reset after each locking cycle.

\begin{algorithm}[H]
\caption{Flowing chart of locking with the ``CI-adaptive" protocol }
\label{alg:locking}
\SetKwInOut{Input}{Input}
\SetKwInOut{Output}{Output}
\SetKwInOut{Initialize}{Initialize}
\SetKwFunction{Actor}{Actor}
\SetKwFunction{Critic}{Critic}
\SetKwFunction{Softmax}{Softmax}
\SetKwFunction{CrossEntropy}{CrossEntropy}

\BlankLine
\Input{minimum interrogation time $T_{min}$;
maximum interrogation time $T_{\rm max}$;
the ``CI-adaptive" protocol iteration steps $n$;
locking steps $n_L$;
time sequence factor $\alpha$;
(K+1) Ensembles $\{N_k\},\{M_k\}$;}
\Initialize{$f=[f_l, f_r]$; $\tilde{\mathcal{P}}(f_c) = 1/ N_0 T_{min}$; $f_{L}^{(0)} = 0$ \;}
[Bayesian Locking loop]:\\
\For{$i = 0$ \KwTo $n_{L}$}{
\BlankLine
[Adaptive Bayesian iteration Loop]:\\
Implement the ``CI-adaptive" protocol as Algorithm.1 and get \{$\mathcal{P}^{(n)}(f_c|\cdots) $, $f_{est}^{(n)}$, $\Delta f_{est}^{(n)}$\}\;
Record the clock transition frequency: $f_c^{(i)} = f_{est}^{(n)}$\;
Update the LO frequency: $f_{L}^{(0)}\leftarrow f_c^{(i)}$\;
Update the interval: $f_l^r \leftarrow f_{est}^{(n)} \pm \Delta f_{\rm est}^{(n)}$\;
Reset the prior: $\tilde{\mathcal{P}}(f_c) = 1/ N_0 T_{min}$.
}
\Output{Clock transition frequencies \{$f_{est}^{(i)}$\} and its overlapping Allan variance.\;}
\BlankLine
\end{algorithm}

\section{ Detection for GHZ-state-based Ramsey interferometry\label{SM:GHZ}}

In this section, we show how to use the practical observable for frequency estimation with GHZ-state-based Ramsey interferometry. 
In general, one can use parity measurement for detection. 
Besides, one can also implement an interaction-based readout and use half-population difference measurement for detection. 
In the following, we show how to use these two protocols for detection.

For an input GHZ state 
\begin{equation}
    \ket{\psi}_{GHZ} =  (\ket{0}^{\otimes N} + \ket{1}^{\otimes N} )/\sqrt{2},
\end{equation}
a relative phase $\phi = \delta T$ is collectively accumulated during the Ramsey interrogation, which is governed by the Hamiltonian 
\begin{equation}
    \hat{H} = \delta \hat{J}_z. 
\end{equation}
Thus, it leads to a $\delta$-dependent output state 
\begin{equation}
    \ket{\psi(\delta)}_{out}=(\ket{0}^{\otimes N} + e^{i N \delta T}\ket{1}^{\otimes N} )/\sqrt{2},
\end{equation}
where $\delta = 2\pi (f_c - f_L)$ is the detuning between the LO frequency $f_L$ and the clock transition frequency $f_c$ between $\ket{0}$ and $\ket{1}$, and $T$ the interrogation time.
Meanwhile, one may vary the accumulated phase by introducing an auxiliary phase $\theta = 2\pi N f_a T$ which can be realized by adding a frequency shift $f_a$ to the local oscillator frequency.

Within the collective spin representation~\cite{nonclassicalstates,biedenharn1984angular}, the eigenvalue of $\hat{J}_z$ is represented by the half population difference $m=(N_\uparrow - N_\downarrow )/2$. 
For an arbitrary quantum state $\ket{\psi} = \sum_{m=-N/2}^{N/2} C_m \ket{N/2,m}$, the expectation of an observable $\hat{O}(\hat{J}_z)$ defined by $\hat{J}_z$ can be described as $\langle \hat{O}(m) \rangle = \sum_{m=-N/2}^{N/2} p_m O(m)  $, where $p_m = |C_m|^2$ is the probability amplitude and $O(m)$ is the eigenvalue of the observable $\hat{O}(\hat{J}_z)$.

For parity measurements $\hat{\Pi} = e^{i\pi \lceil \hat{J}_z \rceil}$, we only need to perform a $\pi/2$ rotation after accumulating the phases $\phi=\delta T$ with a known auxiliary phase $\theta$. The output state for measurement can be expressed as
\begin{equation}
    \label{eq:parity}
    \ket{\psi (\delta)}_{\Pi} = e^{-i \frac{\pi}{2} \hat{J}_y }e^{-i 2\pi f_a T \hat{J}_z} e^{-i\delta T \hat{J}_z} \ket{\psi}_{GHZ}.
\end{equation}
The measurement outcomes for the parity operator $\hat{\Pi}$ are either odd or even, corresponding to $u=-1$ and $u=1$, respectively.
The correspondent probability for obtaining the odd $(u = -1)$ or even $(u = 1)$ parity can be given as 
\begin{eqnarray}
    \label{eq:p_parity}
    \mathcal{L}_{u} &\equiv& \mathcal{L}(u|N,\theta,f_c,f_L,T)\\
    &=& \frac{1}{2}\{1+C\cdot u (-1)^{N} \cos[2\pi NT(f_c-f_L)+\theta]\},\nonumber
\end{eqnarray}
where $C$ is the contrast.  
The probability follows a cosine function and becomes a sine function when $\theta=\pi/2$.
According to Eq.~\eqref{eq:p_parity}, the expectation value of the parity measurement is given by $\langle \hat{\Pi} \rangle = (-1)^N C\cdot \cos[2\pi NT(f_c-f_L)+\theta]$, which revisits the well-known result for the GHZ state.
One can use Eq.~\eqref{eq:p_parity} as the binary likelihood to perform the Bayesian estimation. 

In parallel to parity measurement, we also find that interaction-based readout can be used for GHZ state detection~\cite{huang2018achieving}, which has been shown to be robust to the decoherence induced by spontaneous decay~\cite{kielinski2024ghz}.
After the Ramsey interrogation, one can introduce a one-axis twisting dynamics for interaction-based readout. The output state can be expressed as
\begin{equation}
    \label{eq:sign}
    \ket{\psi (\delta)}_{R} = e^{-i \frac{\pi}{2} \hat{J}_x^2 } e^{-i 2\pi f_a T \hat{J}_z} e^{-i\delta T \hat{J}_z} \ket{\psi}_{GHZ}
\end{equation}
when $N$ is even. 
The ideal probability and the expectation of the half-population difference come out as $p_{\pm \frac{N}{2}} = \frac{1}{2}\{1 \pm (-1)^{ \lceil \frac{N}{2} +1 \rceil }C\cdot \sin[2\pi NT(f_c-f_L)+\theta]\}$ and $\langle \hat{J}_z \rangle  = (-1)^{ \lceil \frac{N}{2} +1 \rceil } \frac{N}{2}C\cdot \sin[2\pi NT(f_c-f_L)+\theta]$.
Besides, one can perform the sign measurement of half-population difference $\hat{S} = Sgn [\hat{J}_z]$ to extract the information of $\delta$. 
The measurement outcomes for the sign measurement $\hat{S}$ are either positive or non-positive with $u=1$ and $u=-1$, respectively. 
The correspondent probability for obtaining the positive $u=+1$ and non-positive $u=-1$ value can be given as
\begin{eqnarray}
    \label{eq:p_sign}
    \mathcal{L}_{u} &\equiv& \mathcal{L}(u|N,\theta,f_c,f_L,T)= \frac{1}{2}\{1 + C\cdot u (-1)^{ \lceil \frac{N}{2} +1 \rceil } \sin[2\pi NT(f_c-f_L)+\theta]\} 
\end{eqnarray}
where $C$ is the contrast.  The probability is a sine function here and would become a cosine function if $\theta=\pi/2$.
According to Eq.~\eqref{eq:p_sign}, the expectation of sign measurement can be given as $\langle \hat{S} \rangle = (-1)^{ \lceil \frac{N}{2} +1\rceil } C \cdot \sin[2\pi NT(f_c-f_L)+\theta]$.
The probability and expectation for $N$ is odd remain the same as stated above if we apply an additional rotation, i.e., $\ket{\psi (\delta)}_{R} = e^{-i \frac{\pi}{2} \hat{J}_y } e^{-i \frac{\pi}{2} \hat{J}_x^2 } e^{-i 2\pi f_a T \hat{J}_z} e^{-i\delta T \hat{J}_z} \ket{\psi}_{GHZ}$.

According to Eqs.~\eqref{eq:p_parity} and \eqref{eq:p_sign}, both protocols have similar probabilities. Thus, we adopt a unified form [Eq.~(1) in the main text] to express the binary likelihood for Bayesian estimation. For parity measurement, $\xi(N)=(-1)^N$, while for interaction-based readout with sign measurement, $\xi(N)=(-1)^{ \lceil \frac{N}{2} +1 \rceil}$. In ideal cases, the contrast in Eqs.~\eqref{eq:p_parity} and \eqref{eq:p_sign} is $C=1$. However, under the influence of dephasing or detection noise, the contrast $C$ will reduce and therefore decrease the precision of the measurement.

\section{Design of the interferometry sequence in ``CI-adaptive'' protocol\label{sec:T_sequence}}

Within the Bayesian framework, we determine the next interrogation time $T_{j+1}$ by constraining the period of the subsequent likelihood function to match the width of the credible interval of the current posterior [as shown in the right panel of Fig.~1~(d) in the main text]:
\begin{equation}
    \label{eq:credible_SM}
    \frac{1}{N_0 T_{j+1}} = 2g\Delta f_{\text{est}}^{(j)},
\end{equation}
which also means $ T_{j+1}=\frac{1}{2g N_0\Delta f_{\text{est}}^{(j)} }   $.
Combining this update rule with the Cram\'{e}r-Rao lower bound under adaptive estimation $\Delta f_{\text{est}}^{(j)} \approx \Delta f_{\text{CRLB}}^{(j)} = 1/(2\pi \kappa \sqrt{N_t} \sqrt{\sum_{i=0}^j T_i^2})$ (where $\kappa = \sqrt{\sum_k M_k N_k^2 / N_t}$ and $N_t = \sum_k M_k N_k$), the optimal interrogation time sequence becomes
\begin{equation}
     T_{j+1}=\frac{\pi \kappa \sqrt{N_t} \sqrt{\sum_{i=0}^j T_i^2}}{g N_0 }
\end{equation}
For simplicity, we take the scaling factor as $\alpha = \frac{\kappa \pi}{g} \frac{\sqrt{N_t}}{N_0}$ and get $T_{j+1}=\alpha \sqrt{\sum_{i=0}^j T_i^2}$.

For two consecutive interrogation times, we have
\begin{align}
    \label{eq:Tj1}
     T_{j+1}^2 =\alpha^{2}\sum_{i=0}^j T_i^2, \\ 
    \label{eq:Tj2}
     T_{j+2}^2 =\alpha^{2}\sum_{i=0}^{j+1} T_i^2 .
\end{align}
By subtracting Eq.~\eqref{eq:Tj1} from Eq.~\eqref{eq:Tj2}, we obtain the following.
\begin{align}
        T_{j+2}^2 - T_{j+1}^2  =\alpha^{2} T_{j+1}^2,
\end{align}
and 
\begin{align}
      \frac{T_{j+2}}{T_{j+1}} = \sqrt{1+\alpha^{2}}.
\end{align}
Therefore, for $j\ge1$, $T_j$ forms a geometric sequence with the initial term $T_1=\alpha T_0$ and the common ratio $\sqrt{1+\alpha^{2}}$, and its general term formula is then given by:
\begin{align}
    T_j=  T_0 \alpha \left(\sqrt{1+\alpha^2}\right)^{j-1}, j\ge 1
\end{align}

Starting from the minimum interrogation time $T_0 = T_{\min}$, and subject to the constraint $T_{\min} \leq T_j \leq T_{\max}$, the predetermined interrogation time is derived as follows:
\begin{equation}
T_j = \min \left\{ \max \left[ T_{\min},\; T_{\min} \alpha \left(\sqrt{1+\alpha^2}\right)^{j-1} \right],\; T_{\max} \right\}.
\end{equation}
This expression, given in Eq. (9) of the main text, defines the interrogation time sequence.

When the interrogation time at each step $j$ remains below the maximum allowed duration ($T_j < T_{\max}$), the total interrogation time $t$ is given by the geometric progression:
\begin{align}
    t = \sum_{j=0}^{n} T_j = T_0 \left[ 1 + \alpha \sum_{j=1}^{n} \left( \sqrt{1+\alpha^2} \right)^{j-1} \right] 
    = T_0 \left[ 1 + \alpha \frac{ \left( \sqrt{1+\alpha^2} \right)^n - 1 }{ \sqrt{1+\alpha^2} - 1 } \right],
    \label{eq:total_time}
\end{align}
where $T_0$ denotes the initial interrogation time and $\alpha$ is a scaling parameter. Solving for $\left( \sqrt{1+\alpha^2} \right)^n$ yields the following:
\begin{align}
    \left( \sqrt{1+\alpha^2} \right)^n = 1 + \frac{t - T_0}{T_0} \beta 
    \label{eq:sqrt_expr}
\end{align}
where $\beta = \sqrt{1+\alpha^{-2}} - \alpha^{-1}$ is defined for notational simplicity.

From the recurrence relation in Eq.~\eqref{eq:Tj1} and Eq.~\eqref{eq:sqrt_expr}, the root sum of squares of the interrogation times is derived as:
\begin{align}
    \sqrt{ \sum_{j=0}^{n} T_j^2 } = \frac{\alpha}{T_{n+1}} = \frac{1}{T_0 \left( \sqrt{1+\alpha^2} \right)^n } = \frac{1}{T_0 + (t - T_0) \beta },
    \label{eq:root_sum_squares}
\end{align}

Thus, the theoretical uncertainty scaling for the frequency estimate, based on the Cramér-Rao lower bound (CRLB), is then expressed as:
\begin{align}
    \Delta f_{\text{CRLB}} = \frac{1}{2\pi \kappa \sqrt{N_t}} \cdot \frac{1}{ \sqrt{ \sum_{j=0}^{n} T_j^2 } } = \frac{1}{2\pi \kappa \sqrt{N_t}} \cdot \frac{1}{ T_0 + (t - T_0) \beta }.
    \label{eq:uncertainty_scaling}
\end{align}

In the case where the interrogation time reaches the maximum allowed duration $T_{\max}$, the uncertainty scaling is bounded by:
\begin{align}
    \Delta f_{\text{CRLB}} = \frac{1}{2\pi \kappa \sqrt{N_t}} \cdot \frac{1}{ \sqrt{ \sum_{j=0}^{n} T_j^2 } } \geq \frac{1}{2\pi \kappa \sqrt{N_t}} \cdot \frac{1}{ \sqrt{ t \cdot T_{\max} } },
    \label{eq:uncertainty_max}
\end{align}
where the inequality arises from the Cauchy-Schwarz inequality applied to the sum of squares.
The lower bound of this uncertainty corresponds precisely to the uncertainty achievable when using only the maximum interrogation time $T_{max}$. This implies that, under ideal conditions, an adaptive protocol can progressively expand the dynamic range while preserving the measurement uncertainty at a level comparable to that of the optimal fixed-interrogation-time strategy.

Combining both cases, the overall theoretical uncertainty scaling versus the total interrogation time $t$ is summarized as follows:
\begin{subnumcases}{\label{eq:delta_ft_SM} \Delta f_{\text{CRLB}} =}
    \dfrac{1}{2\pi \kappa \sqrt{N_t}} \cdot \dfrac{1}{ T_{min} + (t -T_{min}) \beta } & for $T_j \leq T_{\max}$, \\
    \dfrac{1}{2\pi \kappa \sqrt{N_t}} \cdot \dfrac{1}{ \sqrt{t \cdot T_{\max}} } & for $T_j = T_{\max}$,
\end{subnumcases}
where $T_0 \equiv T_{\min}$ denotes the minimum interrogation time. This result corresponds to Eq.~(10) in the main text.

In the case of dephasing~\cite{huelga1997improvement,kielinski2024ghz}, the probability distribution for a single measurement outcome $u$, as given by Eq.~(14) in the main text, is modified to:
\begin{equation}
    \label{eq:dephasing_prob_SM}
    \mathcal{L}_u = \frac{1}{2} \left\{ 1 + e^{-\frac{\gamma}{2} N T} \times \sin[2\pi N T (f_c - f_L) + \theta] \right\}.
\end{equation}
The expectation values derived from this distribution are \( \langle u \rangle = e^{-\frac{\gamma}{2} N T} \sin\phi \) and \( \langle u^2 \rangle = 1 \), where the phase \( \phi = 2\pi N T (f_c - f_L) + \theta \) is introduced for notational simplicity.

Here we take the identical GHZ state as an example for analysis, using the error propagation formula, the uncertainty for $M$ copies in one step is expressed as follows:
\begin{equation}
    \sigma_f = \frac{\Delta u}{ |\partial \langle u \rangle / \partial f_c|} = \frac{1}{2\pi \sqrt{M} N} \cdot \sqrt{ \frac{1 - e^{-\gamma N T} \sin^2\phi }{ T^2 e^{-\gamma N T} \cos^2\phi} }.
    \label{eq:uncertainty_general}
\end{equation}

%
To minimize this uncertainty, the phase is set to \( \phi = 0 \) (which can be achieved by appropriately adjusting the local oscillator frequency \( f_L \) and the phase shift \( \theta \)). Substituting \( \phi = 0 \) into Eq.~\eqref{eq:uncertainty_general} yields the minimum achievable uncertainty:
\begin{equation}
    \sigma_f = \frac{1}{2\pi \sqrt{M} N} \cdot \frac{1}{\sqrt{T^2 e^{-\gamma N T}}}.
    \label{eq:uncertainty_min}
\end{equation}
Similar to Eq.~\eqref{eq:uncertainty_max}, the Cram\'{e}r-Rao lower bound under adaptive estimation then becomes 
\begin{equation}
    \Delta f_{\text{CRLB}} = \frac{1}{2\pi \sqrt{M} N} \cdot \frac{1}{\sqrt{\sum_{i=0}^{j}{T^2 e^{-\gamma N T}} }} \ge \frac{1}{2\pi \sqrt{M} N} \cdot \frac{e^{\gamma N T_{max}/2}}{\sqrt{t T_{max}  }} .
    \label{eq:uncertainty_dephasing_SM}
\end{equation}
The uncertainty in Eq.~\eqref{eq:uncertainty_dephasing_SM} is further minimized with respect to the interrogation time \( T \). The optimal time is found to be \( T_{max}^{opt} = \frac{1}{\gamma N} = \frac{T_2^*}{2N} \), where \( T_2^* = 2 / \gamma \) denotes the coherence time for uncorrelated particles (\( N = 1 \)).
%

\section{Minimum measurement copies and the influences of credible level in ``CI-adaptive" protocol\label{sec:t_table}}

Our assumption of update sequence is based on the lower bound of measurement uncertainty. Therefore, this boundary can only be approached asymptotically if the number of measurements $M$ is sufficiently large.
When $M$ is small, the width $g$ and the rate $\alpha$ are determined by the width of the credible interval according to the t-distribution.
The values of $g$ and $\alpha(\zeta) = \sqrt{M}\pi/g(\zeta)$ under different credible levels $\zeta$ and measurement numbers $M$, are presented in Table.~\ref{tab:t_table}.

When designing the time sequence, the growth rate of the interrogation time is determined by the credible level $\zeta$, which defines the probability that the true frequency lies within the estimated credible interval.
A higher credible level $\zeta$ offers greater stability of the estimation but requires more steps to reach $T_{max}$.  
Conversely, a lower $\zeta$ accelerates convergence to the maximum interrogation time, albeit at the expense of a higher risk of period-skipping errors.

To analyze the influence of the credible level on our protocol, we numerically simulated the protocol with multiple credible levels. We compared the results with the Cram\'{e}r-Rao lower bound (CRLB) in Figs.~2(a) and (c) of the main text and presented the changes in stability with increasing credibility, as shown in Fig.~\ref{fig:mcompare}(a).
Our results demonstrate that stable and good performance can be obtained with a sufficiently high credible level ($\zeta \ge 99.999\%$).
Based on the requirements for time growth ($\alpha \ge 1$) and the credible level ($\zeta \ge 99.999\%$), we found that $M-1 \gtrsim 8$ (e.g., $M \gtrsim 9$) is sufficient to achieve satisfactory performance. This conclusion is verified by numerical simulations, as illustrated in Fig.~\ref{fig:mcompare}(b).
In this analysis, all cases are simulated with the same parameters: $\alpha = 1$, $R = 5000$ repetitions, and throughout the theoretical dynamic range.

\begin{table*}[htp]
	\centering
	\begin{tabular}{ 
		>{\centering\arraybackslash}m{1.5cm} 
		>{\centering\arraybackslash}m{1.5cm}
		*{13}{>{\centering\arraybackslash}m{1.1cm}}
		@{}
	}
		\toprule
		\multicolumn{2}{c}{$M-1$} & 1      & 2      & 3      & 4      & 5      & 6      & 7      & 8      & 9      & 10     & 22     & 1000    & $\cdots$ \\
		\midrule
		\multirow{2}{*}{$\zeta=0.99$} 
		  & $g$      & 63.657 & 9.925  & 5.841  & 4.604  & 4.032  & 3.707  & 3.499  & 3.355  & 3.250  & 3.169   & 2.819  & 2.581   & $\cdots$ \\
		  & $\alpha$ & 0.070  & 0.548  & 1.076  & 1.526  & 1.908  & 2.242  & 2.539  & 2.809  & 3.057  & 3.288   & 5.345  & 38.514  & $\cdots$ \\
		\midrule
		\multirow{2}{*}{$\zeta=0.999$} 
		  & $g$      & 636.619 & 31.599 & 12.924 & 8.610  & 6.869  & 5.959  & 5.408  & 5.041  & 4.781  & 4.587   & 3.792  & 3.300   & $\cdots$ \\
		  & $\alpha$ & 0.007   & 0.172  & 0.486  & 0.816  & 1.120  & 1.395  & 1.643  & 1.870  & 2.078  & 2.272   & 3.973  & 30.117  & $\cdots$ \\
        \midrule
        \multirow{2}{*}{$\zeta=0.9999$} 
          & $g$      & 6366.198 & 99.992 & 28.000 & 15.544 & 11.178 & 9.082  & 7.885  & 7.120  & 6.594  & 6.211   & 4.736 & 3.906 &$\cdots$ \\
          & $\alpha$ & 0.001    & 0.054  & 0.224  & 0.452  & 0.688  & 0.915  & 1.127  & 1.324  & 1.507  & 1.678   & 3.181 & 25.445 &$\cdots$ \\
        \midrule
        \multirow{2}{*}{$\zeta=0.99999$} 
          & $g$      & 63661.977 & 316.225 & 60.397 & 27.772 & 17.897 & 13.555  & 11.225  & 9.783  & 8.827  & 8.150   & 5.693 & 4.4406 &$\cdots$ \\
          & $\alpha$ & $7e-5$    & 0.017  & 0.104  & 0.253  & 0.423  & 0.613  & 0.792  & 0.963  & 1.125  & 1.278   & 2.646 & 22.387 &$\cdots$ \\
        \bottomrule
	\end{tabular}
	\caption{ \textbf{ Credible interval coefficient $g(\zeta)$ and scaling factor $\alpha(\zeta) = \sqrt{M}N\pi/g(\zeta)$ versus degrees of freedom ($M-1$) at fixed credible level $\zeta$.} The $g$ factor determines the credible interval width relative to the posterior standard deviation. Values derived from $t$-distribution statistics with two-tailed significance level $0.01,0.001,0.0001,0.00001$.}
	\label{tab:t_table}
\end{table*}

\begin{figure}[htp]
    \centering
    \includegraphics[width=1.0\linewidth]{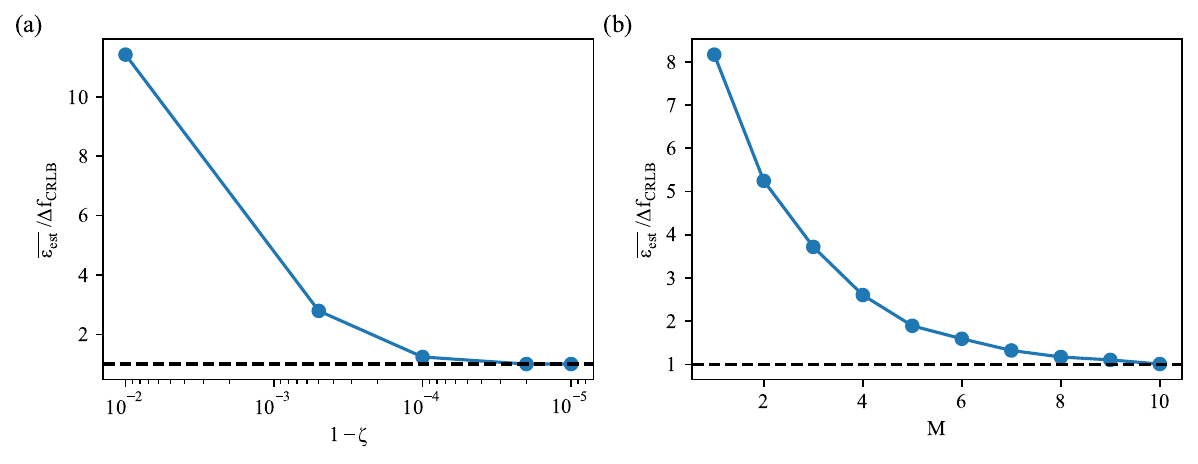}
    \caption{\textbf{Averaged RMSE versus credible level and measurement copies.} (a)  Averaged RMSE versus credible level $\zeta$ when $M=9$. (b) Averaged RMSE versus measurement copy number $M$ with scaling factor $\alpha=1$. All data are averaged over the theoretical dynamic range and $R=5000$ simulations with identical GHZ states ($N=4$).}
    \label{fig:mcompare}
\end{figure}


As illustrated in Figs.~\ref{fig:SM_alpha}~(a)-(c), with sufficiently high credible level (e.g. $\zeta=99.999\%$), the estimates remain stable with minimal period-skipping errors. 
In contrast, as $\zeta$ decreases to $99.9\%$, the probability of errors increases significantly, causing the estimates to converge to incorrect values that differ from the true frequency by integer multiples of the smallest period.
Figs.~\ref{fig:SM_alpha}~(d)-(e) show detailed results from the $544$-th simulation (marked with $\star$ in upper panels), demonstrating how different credible levels affect estimation stability. 
For a high credible level ($\zeta=99.999\%$), even when the initial measurement shows a substantial deviation, the true value remains within the dynamic range of the subsequent estimation (blue shaded area), allowing the estimate to gradually converge to the true frequency (black dashed line). 
However, for lower $\zeta$ values, the credible interval becomes too narrow to contain the true value after initial errors, leading to irreversible period skip errors, as shown in Fig.~\ref{fig:SM_alpha}~(c).

\begin{figure*}[htp]
    \centering
    \includegraphics[width=0.95\linewidth]{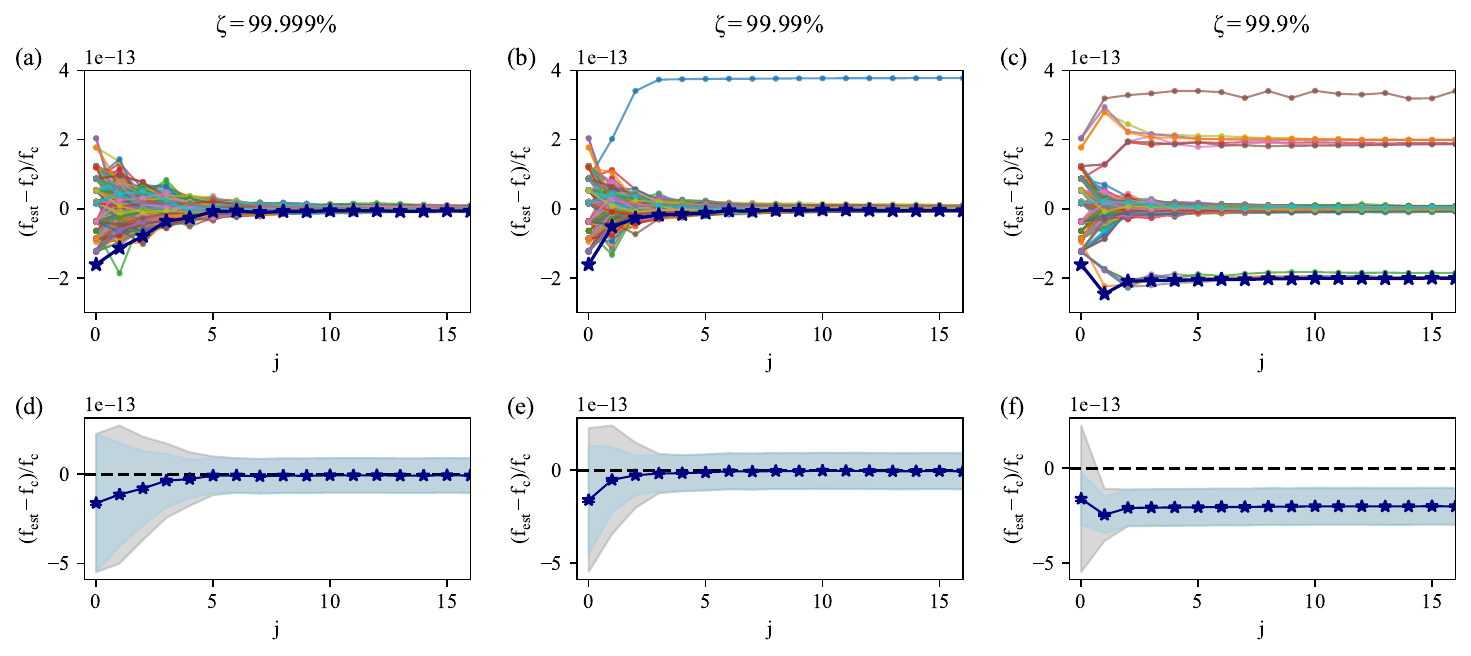}
    \caption{\textbf{Effects of credibility level on estimation stability.} 
    (a--c) Fractional bias as a function of iteration steps for credibility levels $\zeta = 99.999\%$, $99.99\%$, and $99.9\%$, summarizing results from 1000 Monte Carlo simulations. These panels demonstrate the transition from stable convergence to instability as $\zeta$ decreases. 
    (d--e) Detailed trajectory from the $544$-th simulation (marked with $\star$ in upper panel), showing the true frequency (black dashed line), the current estimation uncertainty range (gray shaded area), and the next estimation range based on the credible interval (blue shaded area). 
    All simulations were performed using GHZ states ($N=4$) with identical initial conditions for each random seed.}
    \label{fig:SM_alpha}
\end{figure*}

\section{ Robustness against white noise\label{sec:robustness}}
In this section, we analyze the performance of our protocol in a noisy environment.
We consider two primary noise sources that affect the performance: (i) laser frequency estimation errors caused by white noise, and (ii) signal contrast reduction resulting from dephasing during free evolution and detection processes.
Through comprehensive numerical simulations, we have demonstrated that the ``CI-adaptive" protocol maintains strong estimation performance under moderate noise levels, exhibiting notable resilience against both types of noises.


Usually, the white noise of the laser frequency can be described by a Gaussian distribution around its central frequency~\cite{riehle2003frequency}.
That is, we can incorporate the effects of laser noise as 
\begin{equation}
    f_c^{'} = f_c + f_G,
\end{equation}
where $f_G \sim  \mathcal{G}(0,\sigma_G^2)$ is the Gaussian white noise and $\sigma_G$ is the intensity of the noise.

As an example, we take the same case ($\alpha = 1$) in Fig.~2(a) for comparison.
As illustrated in Fig.~\ref{fig:fig_noise}, although the fractional uncertainty $\Delta\bar{f}_{est}/f_c$ decreases with the total interrogation time $t$, it increases with the noise strength $\sigma_G$. 
Nevertheless, if $\sigma_G \le 7$ Hz, the standard deviation can still outperform the frequentist scheme with the interrogation time fixed as $T_{min}$.
However, when the noise strength is excessively high, the protocol cannot always work well.

\begin{figure}
    \centering
    \includegraphics[width=0.5\linewidth]{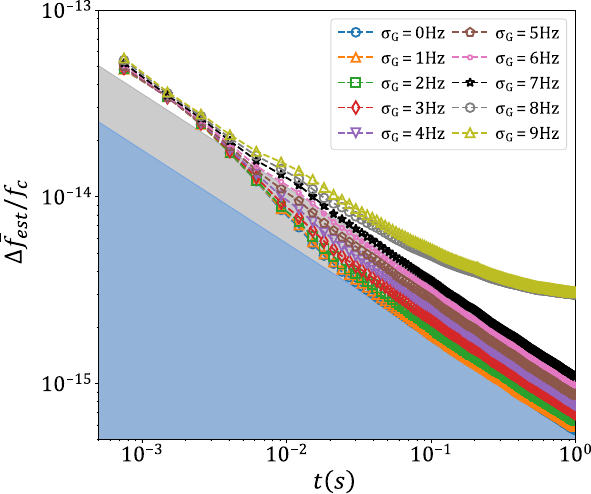}
    \caption{\textbf{Robustness against laser frequency noise.} Fractional uncertainty versus total interrogation time under Gaussian white noise of different intensities $\sigma_G$. Shaded areas show standard deviations from $R=5000$ simulations. Dashed gray line indicates precision surpass SQL. Data generated for $\zeta=99.999\%$ with  GHZ states ($N_0=4$).}
    \label{fig:fig_noise}
\end{figure}

\section{Robustness against detection noise \label{sec:detection}}

In the following, we show how detection noise affects the detection, which is one of the main limitations for realizing practical GHZ-state-based sensors, especially for the parity measurement. 
Despite the fact that both methods can lead to a reduction in contrast, we find that using interaction-based readout with sign measurement is significantly more robust against these noises than conventional parity measurement.

Both parity measurement $\hat{\Pi}$ and sign measurement $\hat{S}$ are based on the half-population difference $\hat J_z$, in which the value of the half-population difference $m$ is obtained and then converted to the corresponding parity or sign value.
For detection noise in realistic measurements, we consider an inefficient detector with Gaussian detection noise~\cite{ma2024phase}, the probability of obtaining the half-population difference $m$ becomes
\begin{equation}
    \label{eq:detection_noise}
    p_m(\sigma_d) = \sum_{n=-N/2}^{n=N/2}{A_n p_n e^{-\frac{(m-n)^2}{2\sigma_d^2} } },
\end{equation}
where $\sigma_d$ is the intensity of the detection noise and $A_n = 1/\sum_{n=-N/2}^{n=N/2}{ p_n e^{-\frac{(m-n)^2}{2\sigma_d^2} }}$ is a normalized factor.
Thus, the contrast under detection noise becomes 
\begin{equation}
    C=\langle{\hat{O}(m)}\rangle _{\sigma_d}/\langle{\hat{O}(m)}\rangle_0,
\end{equation}
where $\langle{\hat{O}(m)}\rangle _{\sigma_d}= \sum_{m=-N/2}^{N/2} O(m) p_m (\sigma_d)$ stands for the reduced expectation of $\hat O$.

To show the influences of detection noise, we compare the contrast $C_{\sigma_d}$ and the metrological gain 
\begin{equation}
    G=20 \rm{log}_{10} \left[ \Delta f_{GHZ}^{opt} (\sigma_d)/ \Delta f_{SCS} \right]
\end{equation}
using three different detection protocols; one is the conventional parity measurement, and the others are interaction-based readout with sign measurement and half-population difference measurement.
Here, metrological gain refers to the optimal precision compared to the corresponding SQL $\Delta f_{SCS}$, and the optimal precision is calculated using the error propagation formula $\Delta f^{opt} = (\Delta \hat{O} / |\partial \langle \hat{O} \rangle /\partial f| )_{f^{opt}}$.

As shown in Fig.~\ref{fig:detectionoise}, it is evident that the parity measurement is significantly affected by detection noise. 
In contrast, the half-population difference and sign measurement demonstrate strong robustness against detection noise, with the sign measurement being the most effective among the three methods.
Consequently, it is feasible to simplify the probability distribution to a binomial distribution by employing interaction-based readout with sign measurement, thereby facilitating Bayesian estimation. This measurement protocol is robust to detection noise and is experimentally friendly for implementation. 

While for detection noise, the parity measurement places high demands on the detector's resolution at the single-particle resolved level.
%
In comparison, the interaction-based readout has lower detector resolution requirements, which enhances its robustness and experimental feasibility under detection noise.
%
If the influences of detection noise are taken into account, the uncertainty and the stability decrease. 
We refer to the experimental results of Ref.~\cite{cao2024multiqubit} and set the parameters to achieve the same contrast in our simulation.
For GHZ states with $N=4$, the ideal case $C=1$ is the same as in Fig.~2.
Under the same conditions, we take the contrast of $C = 0.924$,$C= 0.88(2)$ and $C=0.629$ for comparison.
For the ``CI-adaptive" protocol, the process begins with an interrogation time of $T_{min} = 0.75$ ms, and the initial contrast is $C = 0.92(4)$, which decreases to $C = 0.88(2)$ as the interrogation time $T_j$ increases.
As shown in Fig.~\ref{fig:noises}, under dephasing and detection noise, the ``CI-adaptive" protocol can still accurately extract the clock frequency if $C=0.924$, and $f_{est}$ gradually converges to the true value $f_c$.
When noise is large, more measurements are required to reduce the impact of contrast.

\begin{figure*}[htp]
    \centering
    \includegraphics[width=0.8\linewidth]{fig_SM4.pdf}
    \caption{ \textbf{Robustness of measurement protocols against detection noise.} (a) Contrast versus Gaussian detection noise intensity for conventional parity measurement (solid blue line), interaction-based readout with sign measurement (dashed red line), and half-population difference measurement (dotted green line). Contrast is defined as the ratio between noisy and ideal expectation values. (b) Metrological gain versus detection noise intensity for the same protocols, using identical line styles. Metrological gain is calculated as $20\log_{10}[\Delta f^{\text{opt}}_{\text{GHZ}}(\sigma_d)/\Delta f_{\text{SCS}}]$, where $\Delta f^{\text{opt}}$ denotes optimal frequency uncertainty and $\Delta f_{\text{SCS}}$ is the standard quantum limit for spin coherent states. All data computed for $N=4$ GHZ states.}
    \label{fig:detectionoise}
\end{figure*}

\begin{figure}
    \centering
    \includegraphics[width=0.5\linewidth]{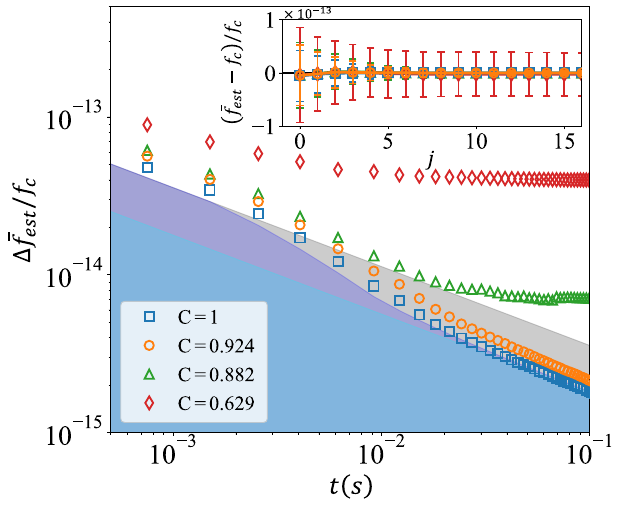}
    \caption{\textbf{Performance under detection noise-induced contrast reduction.} Fractional uncertainty versus total interrogation time for different contrast levels $C$. Shaded areas represent fundamental precision limits. Inset shows bias convergence versus iteration steps. Data from $R=5000$ simulations using cascaded GHZ states ($N_k=\{4,4\}$) with $\alpha=1$.}
    \label{fig:noises}
\end{figure}

\end{widetext}

\end{document}